# Real-Space Tailoring of the Electron-Phonon Coupling in Ultra-Clean Nanotube Mechanical Resonators

A. Benyamini[†,1], A. Hamo[†,1], S. Viola Kusminskiy[2], F. von Oppen[2] and S. Ilani[1,*]

[1] Department of Condensed Matter Physics, Weizmann Institute of Science, Rehovot 76100, Israel.

[2] Dahlem Center for Complex Quantum Systems and Fachbereich Physik, Freie Universität Berlin, 14195 Berlin, Germany.

[†] These authors contributed equally to this work

[*] Correspondence to: shahal.ilani@weizmann.ac.il

**The coupling between electrons and phonons is at the heart of many fundamental phenomena in physics. In nature, this coupling is generally predetermined for both, molecules and solids. Tremendous advances have been made in controlling electrons and phonons in engineered nanosystems, yet, control over the coupling between these degrees of freedom is still widely lacking. Here, we use a new generation of carbon nanotube devices with movable ultra-clean single and double quantum dots embedded in a mechanical resonator to demonstrate the tailoring of the interactions between electronic and mechanical degrees of freedom on the nanoscale. Exploiting this tunable coupling, we directly image the spatial structure of phonon modes and probe their parity in real space. Most interestingly, we demonstrate selective coupling between individual mechanical modes and internal electronic degrees of freedom. Our results open new vistas for engineering bulk quantum phenomena in a controlled nanoscale setting and offer important new tools for entangling the electronic and mechanical degrees of freedom at the quantum level.**

Some of the most well-known phenomena in molecular and solid state physics result from the coupling between electrons and phonons. The resistivity of metals, ferroelectricity, Peierls and Jahn-Teller instabilities as well as BCS superconductivity are different facets of this coupling. In solids, electron-phonon coupling is dictated by the lattice structure and the ensuing electronic bands, leaving little room for tunability. Over the last decades, tremendous advances have been made in the ability to engineer materials on the nanoscale. On the electronic side, artificial atoms – quantum dots – were created, providing extensive control over their electronic spectrum (*1*) which allowed the exploration of a wide variety of phenomena inaccessible in bulk solids. On the mechanical side, a growing variety of engineered systems (*2*, *3*) enabled the study of mechanical phenomena on the nanoscale and brought experiments closer towards controlling the quantum state of mechanical resonators (*4–7*), as well as their coupling to single spins (*8*, *9*), qubits (*5*, *10*) and photons (*11*). These remarkable advances contrast with the still-limited control over the coupling between the electronic and mechanical subsystems, whose tailoring would provide a remarkable toolbox for nano-electro-mechanical systems in the classical and quantum regimes.

Carbon nanotubes (NT) constitute a particularly promising system for tailoring the electron-phonon coupling. Their pristine lattice recently enabled the realization of extremely clean electronic systems (*12–14*), albeit still limited in length and complexity. Moreover, their one-dimensional nature, light mass, and large stiffness enabled the creation of tunable mechanical resonators (*15*, *16*) with high Q-factors (*17*, *18*). Recent pioneering works (*19*, *20*) coupled a NT resonator to a single quantum dot, demonstrating that the mechanical frequency can be strongly affected by a *single* carrier and that the correlated motion of electrons and vibrations can lead to mechanical frequency softening (*21*).

In this work, we explore a new generation of suspended carbon nanotube devices with wide-ranging local control, allowing the formation and manipulation of single and double quantum dots embedded into the mechanical oscillator. We use this unprecedented level of control to shape the interaction between electrons and phonons in real space. The device (Fig. 1A) is created using the nano-assembly technology we recently developed (*22*). It consists of a small band-gap NT suspended between two metallic contacts, 125nm above five electrically independent gates. Above the metallic contacts the NT is hole-doped due to the contacts' workfunction. The suspended part, on the other hand, can be locally doped with either electrons or holes by applying independent DC voltages to the gates, $V_{g1}$ to $V_{g5}$. A negative voltage on all gates dopes the entire NT with holes, effectively creating a continuous 'wire' whose conductance is only weakly gate-dependent. However, when a positive voltage is applied to one of the gates while keeping negative voltages on the others, the NT segment above this gate is doped by electrons, forming a pair of *p-n* junctions that confine a quantum dot above this gate (*22*) (Fig. 1B). As we will show, not only does this dot act as a detector of the local mechanical motion through its charge sensitivity, but more importantly, it provides controlled local coupling between the electronic and mechanical systems, which forms the fundamental building block for this work.

The mechanical vibrations are measured via a standard mixing technique (*15*). A radiofrequency signal with frequency $f$ is applied to an off-center gate (gate 4) and a weaker "probe" signal with frequency $f + \delta f$ is applied to the source electrode (Fig. 1B). The former actuates the mechanical motion and mixes down with the latter via the dependence of the current on source-drain and gate voltages, $\partial^2 I/\partial V_{SD} \partial V_g$, to produce a low-frequency ($\delta f$) current signal measured at the drain. When $f$ is swept through a mechanical resonance, the NT vibration is enhanced, producing a sharp peak in the out-of-phase quadrature of the mixing signal, $M_y$, as well as in the derivative of its in-phase quadrature with respect to $f$, $dM_x/df$ (Fig. 1C, more details in supplementary S1). In this work we use these peaks interchangeably to trace out the mechanical resonance.

Figures 1D and 1E show the measured gate dependence of the first two mechanical modes of the NT resonator. The motion is detected via a quantum dot formed on the resonator at a position of large movement (above gate 3 for the 1st mode and above gate 4 for the 2nd mode, see illustrations). The bottom panels show the measured $M_y$ (colormap) as a function of the voltage on the gate beneath the dot and of the drive frequency, $f$, while the top panels show the simultaneously measured conductance. For both modes the mixing signal is visible whenever the dot is conducting and, being proportional to the derivative of the conductance with respect to gate voltage, it is negative on one side of the Coulomb peak (blue), positive on the other (red), and zero at the peak (white). Similar to previous experiments, we observe that for both modes, the resonance frequencies increase with gate voltage due to the tensioning of the resonator (*15*), that charging by a single electron causes a discrete frequency jump (*19*, *20*) and that the coupling between the electronic and mechanical degrees of freedom causes a sharp softening dip of the resonance frequency concomitant with the Coulomb peak (*19*, *20*). This dynamical coupling will be used here to control the interactions between the mechanical and electronic degrees of freedom.

The first step in exploiting this coupling to tailor the interaction between electrons and phonons is to explore its underlying dynamics. Fundamentally, this coupling results from correlated mechanical and electronic motions: Due to NT vibrations, electrons are pumped between the leads and the quantum dot and their attraction to a biased gate causes a softening of the mechanical restoring force (*21*). This process involves an interesting competition between the vibrational frequency and the electronic tunneling rates, which we study here using a tunable-barrier quantum dot, formed in the resonator over the three central gates and populated with holes (Fig. 2A). The quantum dot can be tuned across the entire range from a closed quantum dot to the open Fabry-Perot-like regime, with individual control over the left and right tunneling rates, $\Gamma_L$ and $\Gamma_R$, by the side gates (1 and 5). We calibrate these rates using transport measurements (Fig. 2B, details in supplementary S2) and measure their independent effect on the mechanical softening. Starting with the symmetric case, $\Gamma_L \approx \Gamma_R$, we observe that the softening of the 1$^{st}$ mechanical mode, $\Delta f_1$, drops with progressive pinching-off of the barriers (Fig. 2C I-III). Plotting the extracted $\Delta f_1$ vs. the total tunneling rate, $\Gamma_L + \Gamma_R$, (Fig. 2D) we observe that the drop commences when the total tunneling rate becomes comparable to the vibrational frequency, $2\pi f_1$. This drop reflects the inability of electrons to follow the mechanical motion and is reproduced by a theoretical calculation (dashed line, details supplementary S6). Interestingly, even when only one barrier is open ($\Gamma_R \gg 2\pi f_1 \gg \Gamma_L$) the softening assumes the maximal value (Fig. 2C IV). This demonstrates that contrary to transport, for which the two barriers add in series, the softening is controlled by the two rates added in parallel, reflecting that the relevant electrons can enter from either lead. Measurements at very large tunneling rates (Supplementary S3) show a similar reduction of $\Delta f_1$, this time due to gradual disappearance of the Coulomb blockade phenomenon (*23*). In between these two drops, there is a wide range of tunneling rates for which the softening remains practically constant (Fig. 2D and Supp. Fig. S4). In this regime electrons enter sufficiently fast to establish electrostatic equilibrium at all times but not so fast as to broaden the Coulomb blockade peaks beyond their thermal broadening. The different regimes of the coupling dynamics, demonstrated above, will be utilized below to explore the different facets of this coupling.

A key feature that allows us to tailor the coupling between electrons and phonons is the control over the real-space confinement of the electrons. With five gates we can localize a quantum dot at five different locations along the tube, and explore how its position affects the coupling. To eliminate spurious position-dependence effects, we ensure that all the parameters that are relevant for softening are similar for dots formed at the different locations. Specifically, all dots have the same number of electrons, similar charging energies and similar gate couplings (supplementary S4), and their tunneling rates are chosen well within the range where they do not affect the softening, as explained above. Interestingly however, when measuring the softening of the 1$^{st}$ phonon mode with dots at the five locations (Figs. 3A-E, illustration in each panel) we observe that it depends strongly on position: It is weak near the contacts and increases continuously until reaching a maximum at the center of the resonator. Figures 3F-J show similar measurements for the 2$^{nd}$ mechanical mode. Again we observe strong position dependence, however, in contrast to the 1$^{st}$ mode the softening is practically zero at the center, and has its maxima above gates 2 and 4. Plotting the extracted softening for both modes, $\Delta f_1$ and $\Delta f_2$, as a function of the spatial position of the quantum dot (Figs. 3K and 3L) we find that they nicely follow the spatial displacement profile of the corresponding phonon modes (plotted as lines). The coupling thus provides a direct imaging of the shapes of the phononic modes in real space.

We can understand why the spatial dependence of the coupling follows the shapes of the phononic modes from a simple physical picture of the electrostatic forces. In supplementary S8 we calculate the local force acting on the NT resonator due to single-electron charging of a localized quantum dot. We show that this force can be viewed as being due to an effective "electronic spring" with a negative spring constant, "attached" at the position of the dot (illustrated in Fig 3K). The spring constant does not depend on the position of the dot along the tube, but the shift in the frequency of the combined system does: If the spring is connected at a node of the phononic mode it has no effect on its dynamics, whereas if it is connected at a location of large vibrational amplitude, it has a strong effect. Indeed, perturbation theory shows (Supplementary S8) that the frequency shift is proportional to the amplitude squared of the bare phonon mode at the location of the local spring, in agreement with our observation. Beyond providing direct imaging of the phonon modes, shown in (*24*), the above measurements demonstrate that by moving the quantum dot in real space it is possible to turn on and off its coupling to selected phonon modes, thereby creating controllable coupling between these degrees of freedom.

Is it possible to tailor the coupling between electrons and phonons without moving the electronic confinement potential? To achieve this we increase the spatial complexity of the electronic system and form, for the first time in a NT resonator, a double quantum dot. The added degrees of freedom allow us to generate specific couplings between individual electronic and phononic modes, with some similarities to the coupling with photons reported recently (*25*, *26*). The double dot is formed above gates 2 and 4, which also act as the corresponding plunger gates, and its left, right and center tunnel barriers are controlled by the remaining three gates. The measured double-dot conductance as a function of the left and right plunger gate voltages (Fig. 4A) shows an extremely clean charge stability diagram, down to the single-electron limit. For the experiment we zoom onto a symmetric charge transition vertex (Fig. 4B) and study the two complementary facets of the electron-phonon coupling. We start by exploring the effects of the phonons on the electrons, through which we demonstrate the mode selectivity of this coupling. We then demonstrate the complementary effect of internal double-dot electronic modes on the phonons.

The coupling effects on the electrons are imprinted in the magnitude and sign of the mixing signal, as opposed to the frequency shift discussed so far. The former is isolated in Fig. 4C which plots the mixing signal $M_y$, measured for the 1st mechanical mode over the same voltage range as in Fig. 4B, but with the frequency shifts integrated out (see caption). A similar measurement for the 2nd mechanical mode is shown in Fig. 4D. Curiously, we see that the patterns of negative and positive mechanical mixing signal (blue/red) are substantially different for these two modes. This difference reveals the distinct way that different phononic modes act on the electrons: In the 1st mode, the two dots are moving in phase, getting closer and further away together from the plunger gates (inset, Fig. 4C). In the 2nd mode, the dots move out-of-phase, with one approaching while the other receding from the gates (inset, Fig. 4D). This different mechanical motion translates into different electrical gating: For equal DC voltages on the plunger gates, the 1st mode gates the double dot along the common-mode voltage direction (vector $V$ in Fig. 4B) whereas the 2nd mode gates it along the detuning direction (vector $\varepsilon$ in Fig. 4B). Each phonon mode can thus be mapped onto an effective "gating vector" in voltage space, and correspondingly, its mixing signal should be the derivative of the conductance along the direction of this vector. By taking the numerical derivative of the conductance in Fig. 4B along the $V$ and $\varepsilon$ directions (Figs. 4E and 4F) we indeed observe excellent agreement with the measured mixing signals of the 1st and 2nd modes, respectively. The above measurement demonstrates two important aspects: First, it shows that it is possible to use the electrons to directly probe the real-space parity of the phonons. But furthermore, it demonstrates that each phonon mode has a characteristic action on the electrons, captured by its "gating vector", providing a powerful tool for tailoring selective coupling between these degrees of freedom.

So far we have demonstrated various aspects of tailored coupling, but this coupling was to random electrons tunneling in and out from the leads. If one can couple, however, the phonons to *internal* electrons in an isolated system, one can imagine realizing clean systems decoupled from random electronic degrees of freedom. To study this possibility we isolate the double-dot from the leads by symmetrically pinching-off its side barriers while maintaining a large internal tunneling rate, $\Gamma_C \gg 2\pi f_i$ ($i = 1,2$). Fig. 4G shows the softening of the 1st and 2nd modes measured in this regime along their effective gating directions ($V$ and $\varepsilon$ respectively) and plotted as a function of a normalized tunneling rate to the leads, $\gamma = (\Gamma_L + \Gamma_R)/2\pi f_i$. As the double dot is disconnected from the leads by lowering $\gamma$, we observe a clear quenching of the 1$^{st}$ mode softening, in complete analogy with the observation for the single dot case (Fig. 2). Remarkably, however, the detachment from the leads leaves the softening of the 2$^{nd}$ mode essentially unaffected. This intriguing observation can be understood by realizing that a double dot has an internal electronic degree of freedom, involving the transfer of charge between the dots, which provides the correlated electron flow that induces the softening (illustration, Fig. 4G). This degree of freedom does not couple to the common-mode gating of the 1$^{st}$ mode but directly couples to the detuning gating of the 2$^{nd}$ mode and thus softens only the latter, an effect that is nicely captured theoretically (Supplementary S9). The above measurements thus clearly show that it possible to couple phonons to *internal* electronic degrees of freedom and that this coupling is selective.

The demonstration of tailored coupling between internal electronic and phononic degrees of freedom opens a wide range of new possibilities. One example pertains to the coupling of phonons to solid-state qubits, which use the singlet and triplet states of two electrons in a double quantum dot as their basis (*27*, *28*). Due to the Pauli blockade, an electron can shift between the dots only in the singlet state but not in the triplet state, and thus the 2$^{nd}$ phonon frequency will dynamically couple only to the former. This selective coupling thus provides a tantalizing new route for transferring quantum entanglement from the electronic to the mechanical subsystems, or even between distant qubits in a multi-site lattice via a phonon "bus" (*29*, *30*). Generalizing the physics demonstrated here in double-dots to multi-site lattices that are now well within reach (*22*) would enable an even broader class of experiments that explore bulk electron-phonon phenomena, such as ferroelectricity, Peierls and Jahn-Teller instabilities, or superconductivity, in an engineered nanoscale setting. Analogous to the richness of quantum dot physics, made possible by the extensive control over their electronic properties, the ability to tailor the dynamics, spatial structure and selectivity of the coupling between electrons and phonons, demonstrated here, will enable studying these phenomena in new regimes unattainable in bulk systems, opening new frontiers for fundamental experiments in condensed matter physics.

**References and Notes:**

1. M. A. Kastner, Artificial Atoms, *Physics Today* **46**, 24 (1993).

2. K. L. Ekinci, M. L. Roukes, Nanoelectromechanical systems, *Review of Scientific Instruments* **76**, 061101 (2005).

3. M. Poot, H. S. J. van der Zant, Mechanical systems in the quantum regime, *Physics Reports* **511**, 273–335 (2012).

4. M. D. LaHaye, O. Buu, B. Camarota, K. C. Schwab, Approaching the quantum limit of a nanomechanical resonator., *Science* **304**, 74–7 (2004).

5. A. D. O'Connell *et al.*, Quantum ground state and single-phonon control of a mechanical resonator., *Nature* **464**, 697–703 (2010).

6. J. D. Teufel *et al.*, Sideband cooling of micromechanical motion to the quantum ground state., *Nature* **475**, 359–363 (2011).

7. J. Chan *et al.*, Laser cooling of a nanomechanical oscillator into its quantum ground state, *Nature* **478**, 89–92 (2011).

8. O. Arcizet *et al.*, A single nitrogen-vacancy defect coupled to a nanomechanical oscillator, *Nature Physics* **7**, 879–883 (2011).

9. S. Kolkowitz *et al.*, Coherent sensing of a mechanical resonator with a single-spin qubit., *Science* **335**, 1603–6 (2012).

10. M. D. LaHaye, J. Suh, P. M. Echternach, K. C. Schwab, M. L. Roukes, Nanomechanical measurements of a superconducting qubit., *Nature* **459**, 960–4 (2009).

11. M. Aspelmeyer, P. Meystre, K. Schwab, Quantum optomechanics, *Physics Today* **65**, 29 (2012).

12. J. Cao, Q. Wang, H. Dai, Electron transport in very clean, as-grown suspended carbon nanotubes., *Nature materials* **4**, 745–9 (2005).

13. F. Kuemmeth, S. Ilani, D. C. Ralph, P. L. McEuen, Coupling of spin and orbital motion of electrons in carbon nanotubes., *Nature* **452**, 448–52 (2008).

14. G. A. Steele, G. Gotz, L. P. Kouwenhoven, Tunable few-electron double quantum dots and Klein tunnelling in ultraclean carbon nanotubes., *Nature nanotechnology* **4**, 363–7 (2009).

15. V. Sazonova *et al.*, A tunable carbon nanotube electromechanical oscillator., *Nature* **431**, 284–7 (2004).

16. R. Leturcq *et al.*, Franck–Condon blockade in suspended carbon nanotube quantum dots, *Nature Physics* **5**, 327–331 (2009).

17. A. K. Hüttel *et al.*, Carbon nanotubes as ultrahigh quality factor mechanical resonators., *Nano letters* **9**, 2547–52 (2009).

18. M. Ganzhorn, W. Wernsdorfer, Dynamics and Dissipation Induced by Single-Electron Tunneling in Carbon Nanotube Nanoelectromechanical Systems, *Physical Review Letters* **108**, 175502 (2012).

19. B. Lassagne *et al.*, Coupling mechanics to charge transport in carbon nanotube mechanical resonators., *Science* **325**, 1107–10 (2009).

20. G. A. Steele *et al.*, Strong coupling between single-electron tunneling and nanomechanical motion., *Science* **325**, 1103–7 (2009).

21. M. T. Woodside, P. L. McEuen, Scanned probe imaging of single-electron charge states in nanotube quantum dots., *Science* **296**, 1098–101 (2002).

22. J. Waissman *et al.*, Electronically-Pristine and Locally-Tunable One-Dimensional Systems Created in Carbon Nanotubes Using Nano-Assembly, *Preprint at arXiv:1302.2921* .

23. H. B. Meerwaldt *et al.*, Probing the charge of a quantum dot with a nanomechanical resonator, *Physical Review B* **86**, 115454 (2012).

24. D. Garcia-Sanchez *et al.*, Mechanical Detection of Carbon Nanotube Resonator Vibrations, *Physical Review Letters* **99**, 085501 (2007).

25. T. Frey *et al.*, Dipole Coupling of a Double Quantum Dot to a Microwave Resonator, *Physical Review Letters* **108**, 046807 (2012).

26. K. D. Petersson *et al.*, Circuit quantum electrodynamics with a spin qubit., *Nature* **490**, 380–3 (2012).

27. R. Hanson, L. P. Kouwenhoven, J. R. Petta, S. Tarucha, L. M. K. Vandersypen, Spins in few-electron quantum dots, *Reviews of Modern Physics* **79**, 1217–1265 (2007).

28. F. Pei, E. A. Laird, G. A. Steele, L. P. Kouwenhoven, Valley-spin blockade and spin resonance in carbon nanotubes., *Nature nanotechnology* **7**, 630–4 (2012).

29. J. Cirac, P. Zoller, Quantum computations with cold trapped ions, *Physical Review Letters* **74**, 4091–4094 (1995).

30. Q. Turchette *et al.*, Deterministic Entanglement of Two Trapped Ions, *Physical Review Letters* **81**, 3631–3634 (1998).

**Acknowledgements:** We acknowledge O. Auslaender, E. Berg, F. Kuemmeth, P. L. McEuen, A. Shnirman and A. Yacoby for useful discussions and comments to the manuscript, and D. Mahalu for the e-beam writing. S.I. acknowledges the financial support by the ISF Legacy Heritage foundation, the Bi-National science foundation (BSF), the Minerva foundation, the ERC starters grant, the Marie Curie People grant (IRG), and the Alon fellowship. S.I. is incumbent of the William Z. and Eda Bess Novick career development chair. FvO acknowledges support through SPP 1459, SFB 658 as well as a Helmholtz Virtual Institute.

**Fig. 1: A carbon nanotube mechanical resonator coupled to localized ultra-clean quantum dots.** (A) Scanning electron micrograph of a device similar to the one measured, scale bar 100nm. (B) Measurement layout: DC gate voltages, $V_{g1}$ to $V_{g5}$, locally dope the NT with electrons (red) or holes (blue). Mechanical motion is actuated by an RF signal on gate 4 (frequency $f$) and is detected as a low frequency signal ($\delta f$) in the drain by down-mixing with a weak probe signal of frequency $f + \delta f$ applied at the source. (C) Various mixing signal components measured as a function of the drive frequency, $f$: In-phase quadrature, $M_x$ (blue), out-of-phase quadrature, $M_y$ (green), and the derivative $dM_x/df$ (purple). (D) Top: Conductance, $G$, of a dot above gate 3 as a function of $V_{g3}$. Bottom: Corresponding mixing signal, $M_y$ (colormap), measured for the 1$^{st}$ mechanical mode, as a function of $V_{g3}$ and $f$. Dashed gray line is a fit to a theory including only the static electron-phonon coupling, capturing the frequency step across a Coulomb blockade peak. Dashed black line includes also the dynamical coupling (supplementary S1). Their difference at the center of the Coulomb peak, $\Delta f_1$, gives the dynamic frequency softening. (E) Similar measurement for the 2$^{nd}$ mechanical mode with a dot above gate 4. All measurements in this paper are done at an electron temperature of $T = 16K$ as determined from the Coulomb peaks in the conductance.

**Fig. 2: Dependence of dynamical electro-mechanical coupling on the electron tunneling rate.** (A) Schematics: A quantum dot of holes is created above gates 2-4. Its left and right barriers tunneling rates, $\Gamma_L$ and $\Gamma_R$, are controlled by $V_{g1}$ and $V_{g5}$. (B) Measurement of the peak conductance at the Coulomb blockade transition from 5 to 6 holes, $G_{peak}$ (colormap), as a function of $V_{g1}$ and $V_{g5}$, from which $\Gamma_L$ and $\Gamma_R$ are independently extracted (see supplementary S2 for details). (C) Mixing signal of the 1$^{st}$ mode, $dM_x/df$ (colormap), measured as a function of $V_{g3}$ and $f$ for symmetric tunneling rates, $\Gamma_L \approx \Gamma_R$ (subpanels I-III), and for asymmetric rates, $\Gamma_L \ll \Gamma_R$ (subpanel IV). (D) Extracted softening, $\Delta f_1$, as a function of the total tunneling rate, $\Gamma = \Gamma_L + \Gamma_R$. The dashed line is a fit to the theory in Supplementary S6. Dotted vertical line marks the angular frequency of the mechanical mode, $2\pi f_1$.

**Fig. 3: Spatial dependence of the electro-mechanical coupling, and a direct imaging of the phonon modes.** (A)-(E) Mixing signals of the 1$^{st}$ phonon mode, $dM_x/df$ (colormap), measured with quantum dots formed above each of the five gates (see illustrations). Vertical bars are $0.5\ MHz$. The softening shows a clear dependence on the quantum-dot position. (F)-(J) Similar measurements for the 2$^{nd}$ mode, yielding a different position dependence. (K) and (L) The softening of the 1$^{st}$ and 2$^{nd}$ modes, $\Delta f_1$ and $\Delta f_2$, extracted from panels (A)-(J) and plotted as a function of the positions of the quantum dots, taken from the lithographic positions of the gates. Lines: Calculated amplitude squared of the corresponding phonon mode vs. position in the beam (solid) and string (dashed) limits (Supplementary S4). Inset to panel (K): Effective mechanical model: The force exerted on the NT by electrons flowing through a localized dot is equivalent to a spring with a negative spring constant attached at the dot's position.

**Fig. 4: Tailored selective coupling between phonon modes and *internal* electronic degrees of freedom in a double quantum dot.** (A) Conductance, $G$ (colormap), of a double-dot defined above gates 2 and 4 measured as a function $V_{g2}$ and $V_{g4}$. The number of electrons in both dots are labeled by $(n, m)$. (B) Zoom-in on the (3,4) to (4,3) transition. Common mode and detuning gating directions are labeled $V$ and $\varepsilon$. (C) 1$^{st}$ mode mixing, $M_y$ (colormap), measured over the same voltage window as in panel B. (D) Same for the 2$^{nd}$ mode. In both cases we subtract the electronic mixing signal measured away from the mechanical resonance and integrate over a small frequency window to project out the frequency shifts due to softening (Supplementary S1). Insets: mapping the mechanical motion onto effective electrical gating (see text) (E) and (F) Numerical derivatives of the conductance in panel B along the $V$ and $\varepsilon$ directions, respectively, showing good agreement with the measured mechanical mixing (panels C and D). (G) Coupling to an internal electronic mode in a double dot. As a function of the normalized tunneling rate to the leads, $\gamma = (\Gamma_L + \Gamma_R)/2\pi f_i$ ($i = 1,2$), we plot the measured softening of the 1$^{st}$ and 2$^{nd}$ modes, normalized by their asymptotic value at large $\gamma$, $\Delta f_i/\Delta f_i^{max}$. The central barrier is kept fixed with $\Gamma_C \gg 2\pi f_i$. Dashed line: Fit to the single-dot theory as in Fig. 2, capturing the 1$^{st}$ mode softening roll-off, but not the $\gamma$-independent softening of the 2$^{nd}$ mode. Top insets sketch the relevant electron tunneling mechanisms for connected

(right) and isolated (left) double-dots. Measurements in this panel are done around the (6,7) to (7,6) hole vertex (Supplementary S5).

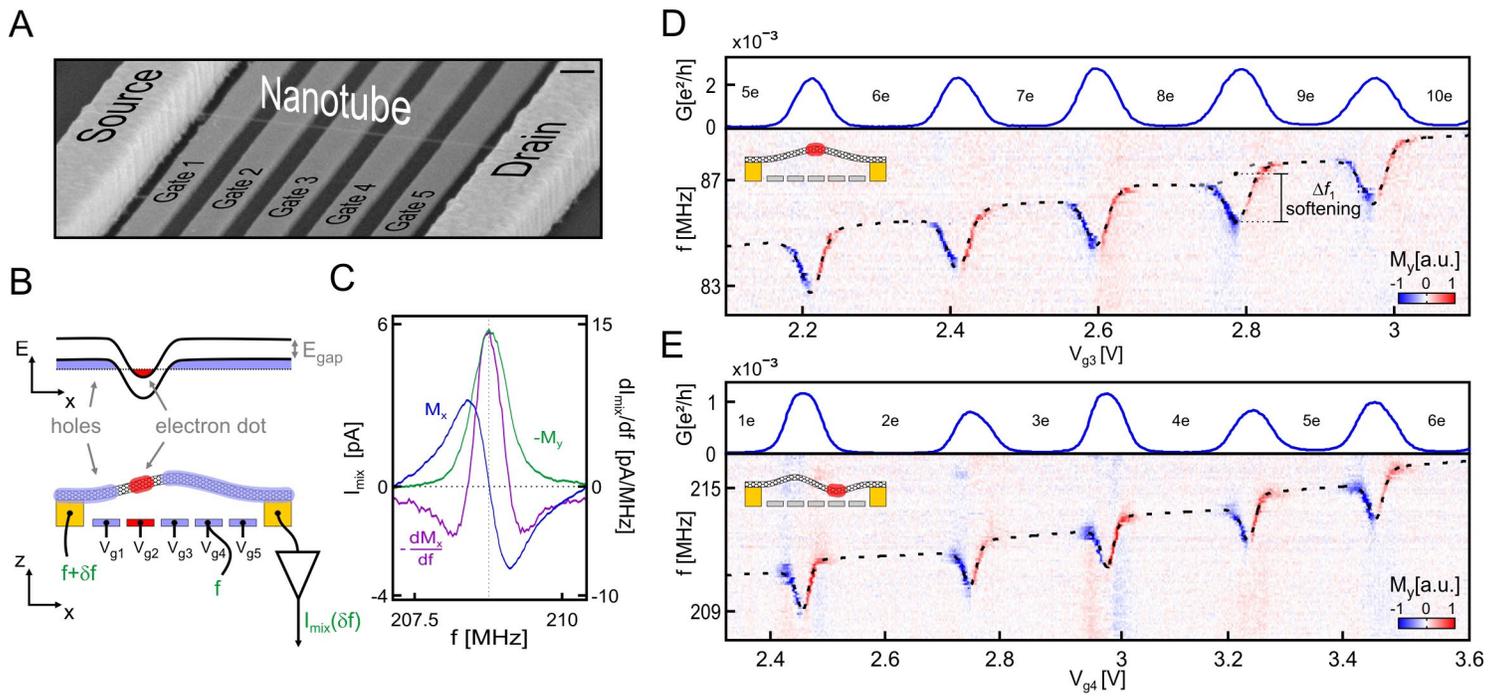

figure 1

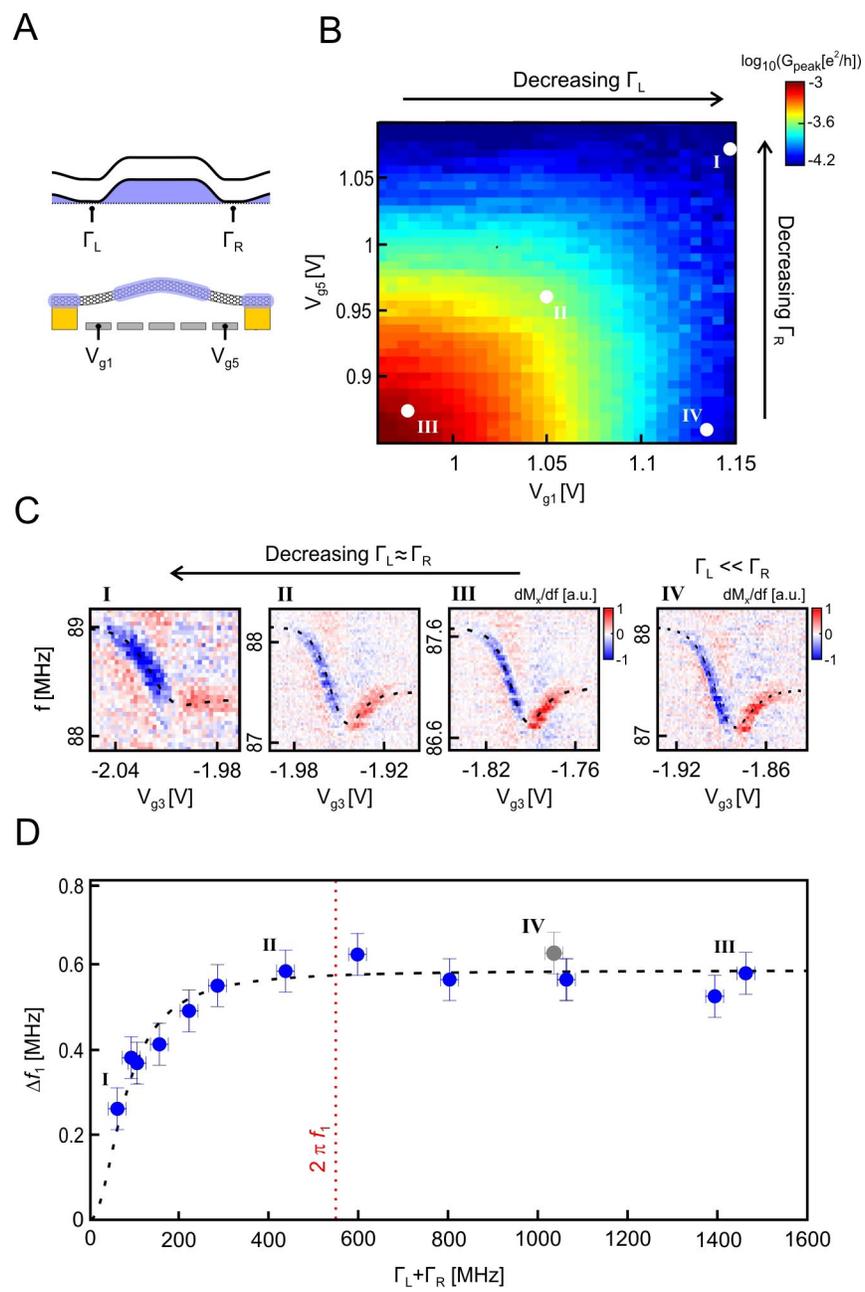

figure 2

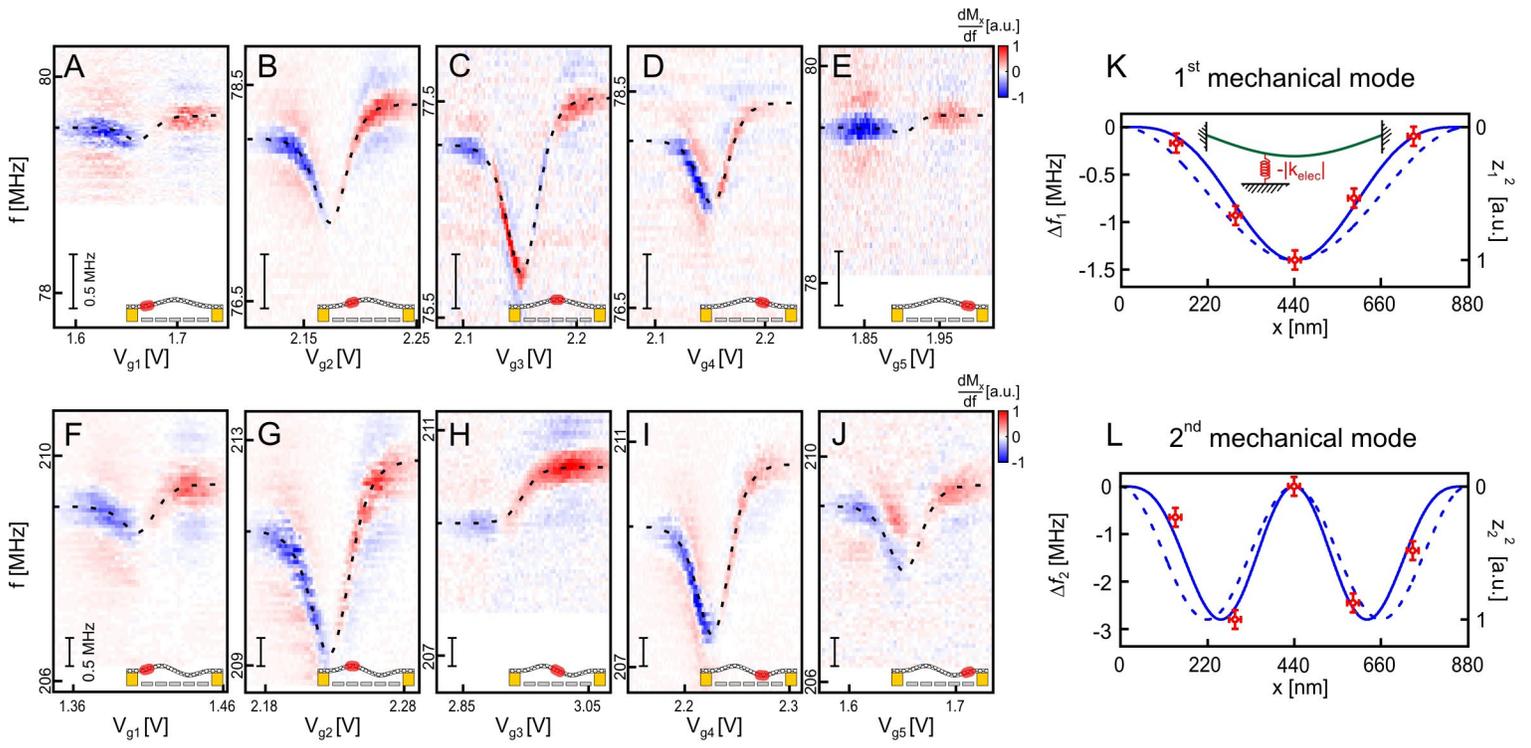

figure 3

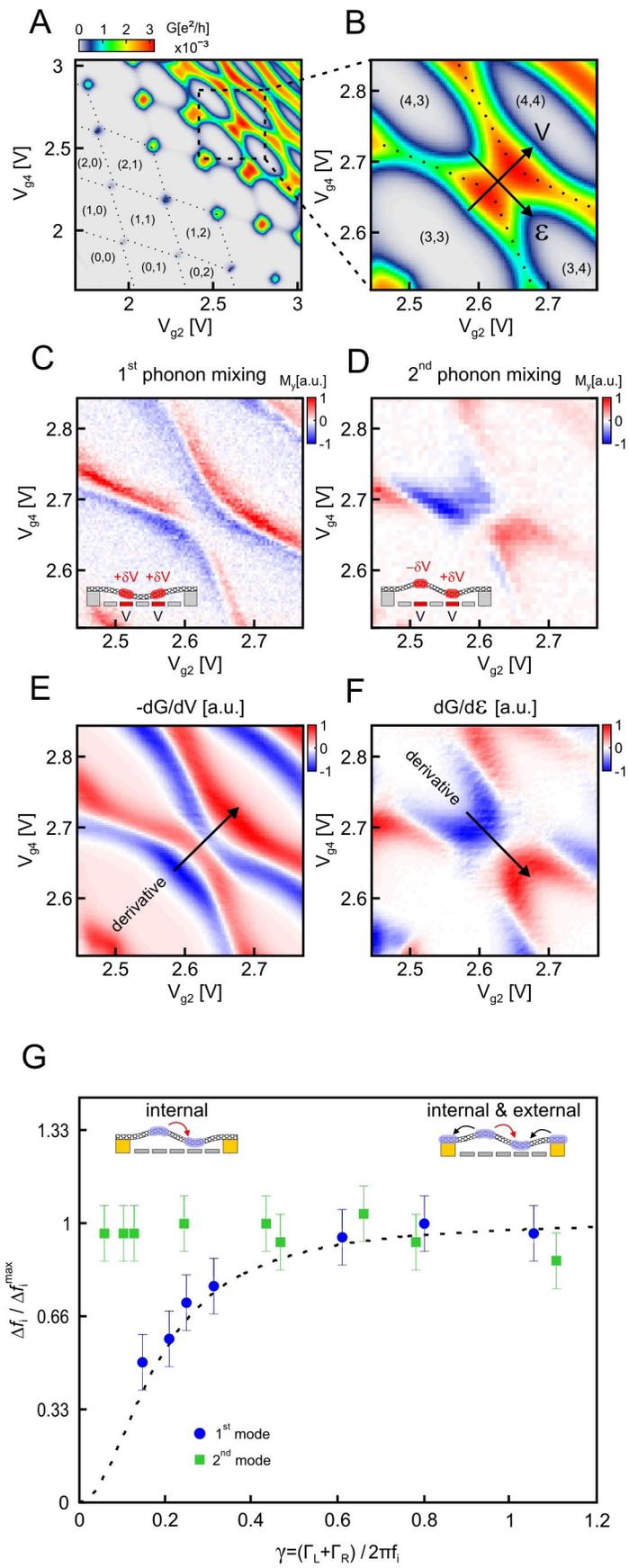

figure 4

# Supplementary Materials

## Real-Space Tailoring of the Electron-Phonon Coupling in Ultra-Clean Nanotube Mechanical Resonators

A. Benyamini*[1], A. Hamo*[1], S. Viola Kusminskiy[2], F. von Oppen[2] and S. Ilani[1]

S1. **Identifying the mechanical resonance in the two quadratures of the mixing signal**
S2. **Determining the tunneling rates from conductance measurements**
S3. **The fate of softening at large tunneling rates: the open quantum dot regime.**
S4. **Comparison of the position-dependent softening and vibration mode profiles**
S5. **Details of the mechanical softening for a double dot**
S6. **Theory: Rate dependence of softening**
S7. **Theory: Generalization to multiple gates**
S8. **Theory: Spatial dependence of the softening**
S9. **Theory: Softening with a double-dot**

S1. **Identifying the mechanical resonance in the two quadratures of the mixing signal**

In this section we explain in more detail the mixing scheme used to measure the mechanical resonances, describe the measured quantities, and explain how the resonance frequency is determined from these measurements.

The NT motion is actuated using an oscillating potential of frequency $f$ applied to a gate, $\delta V_g \sim \cos(2\pi f t)$. This potential generates an electrostatic force that drives the NT motion, and its application to an off-center rather than a center gate allows actuating also high-lying mechanical modes, independent of them being symmetric or anti-symmetric in real space.

The mechanical motion is detected via measurements of the current through the NT, $I$ (1–6). The latter depends on the "control charge" induced on the NT by a gate, $Q_c = C_g V_g$ ($C_g$ and $V_g$ are the gate capacitance and potential with respect to the NT), a dependence which is especially strong in the Coulomb blockade regime. The effect of an oscillating gate potential on $Q_c$ (and consequently on $I$) has two independent contributions: $\delta Q_c = C_g \delta V_g + \delta C_g V_g$. The first term is of pure electronic origin, coming from direct electrostatic gating. The second term has a mechanical origin: A NT oscillation with an amplitude $\delta z$ causes an oscillating gate capacitance $\delta C_g = \frac{dC_g}{dz}\delta z$, which leads to an oscillatory $\delta Q_c$. Since the current depends on the capacitance only through the control charge, one can map the effect of the mechanical motion into an effective "mechanical gating": $\delta V_g^{mech} = \frac{V_g}{C_g}\frac{dC_g}{dz}\delta z$. When $f$ is far from a mechanical resonance, the mechanical motion is small and the electronic contribution dominates. Near a resonance $\delta z$ is enhanced significantly and the mechanical term becomes dominant.

The above effects produce oscillations of the NT current at a frequency $f$, which is typically too high (~100MHz) to be detected directly due to the RC time constant at the output of the NT. To detect this signal we therefore downmix it, via a non-linearity in the NT transport, with a weak "probe" signal on the source contact, $\delta V_s \sim \cos[2\pi(f + \delta f)t]$, to produce a low frequency signal ($\delta f = 1kHz$) at the drain: $\delta I \sim \frac{d^2 I}{dV_g dV_s} \cos(2\pi \delta f t)$. In the formula above the gate oscillations could be either due to the electronic gating or the effective mechanical gating. In this gate-source mixing scheme the mixing signal is thus proportional to $\frac{d^2 I}{dV_g dV_s} = \frac{dG}{dV_g}$ ($G$ is the conductance).

Figure S1 shows the mixing signal measured across a Coulomb blockade peak in a quantum dot formed above gate 4, plotted as a function of the DC voltage on this gate, $V_{g4}$, and the drive frequency, $f$. Panel A shows the in-phase quadrature of this signal, $M_x$, panel B shows the out-of-phase quadrature, $M_y$, and panel C shows the numerical derivative of $M_x$ with respect to $f$. In the $M_x$ panel we subtracted the electronic contribution measured ~2MHz away from the mechanical resonance. The top insets show the simultaneously-measured quasi-DC conductance, exhibiting a Coulomb peak that corresponds to a transition from zero to one electron occupation in the dot.

Notably, $M_x$ flips sign (red to blue) both as a function of frequency and as a function of gate voltage. The sign flip as a function of frequency occurs at the mechanical resonance frequency and results from the mechanical motion switching from being in-phase with the electrical actuation below this frequency, to being out-of-phase with it above this frequency. This sharp phase rotation at resonance also leads to a finite signal in the out-of-phase quadrature, $M_y$, peaking at the resonance frequency (Fig. S1b). A similar peak is also obtained in the derivative $\frac{dM_x}{df}$ (Fig. S1c). In the main text we use the sharp peaks in either of the last two quantities as a clear signature for the resonance frequency.

The sign flip as a function of gate voltage occurs very close to the peak in the conductance. This flip reflects the fact that our gate-source mixing scheme probes predominantly a specific non-linear term in the NT transport, the transconductance, $dG/dV_g$, which changes sign at the Coulomb peak. The effect is also clearly visible in the out-of-phase quadrature, $M_y$, (Fig. S1b), which is negative (blue) on one side of the Coulomb peak, and positive (red) on the other side of the peak.

The dashed line in all panels is a fit for the following theoretical expression, derived in section S6:

$$f = f_0 + \Delta f_{static} \cdot F'(\nu) - \Delta f_{dynamic} \cdot \cosh^{-2}(\nu), \tag{S1}$$

where $\nu = \frac{\alpha(V_g - V_0)}{2k_B T}$, $\alpha = \frac{C_g}{C_\Sigma}$, $C_\Sigma$ is the dot's self-capacitance, $k_B$ is the Boltzmann constant, $V_0$ is the position in voltage of the Coulomb peak, $T$ is the temperature, and $F'$ is the derivative of the Fermi function. The first term gives the 'classical' frequency of the NT resonator due to its flexural rigidity and mechanical tension induced by the gate voltage. The second term gives the static frequency shift leading to the step increase due to one electron addition, and the last term describes the softening due to dynamic coupling at the charge transition. For the static term we assumed $V_g \gg q/C_\Sigma$. In figure 1 and 4G of the main text, this formula is generalized to the case of multiple charge transitions, which are well separated in a single dot and are slightly overlapping in a double dot when going along the common voltage direction.

In figure 4 of the main text we measure $M_y$ as a function of two gate voltages, $V_{g2}$ and $V_{g4}$, and exploit the fact that it is proportional to the gate derivative of the conductance, $dG/dV_g$, to determine the effective 'mechanical gating' direction of the 1$^{st}$ and 2$^{nd}$ mechanical modes. We note that as a function of the gate voltages the resonance frequency is also changing, mostly due to the softening, as is shown for example in Fig. S1b. To eliminate this effect in figure 4 and display only the magnitude of $M_y$ we integrate this quantity over a small frequency window around the mechanical resonance, thus capturing the integrated strength of the peak irrespective of its frequency shifts.

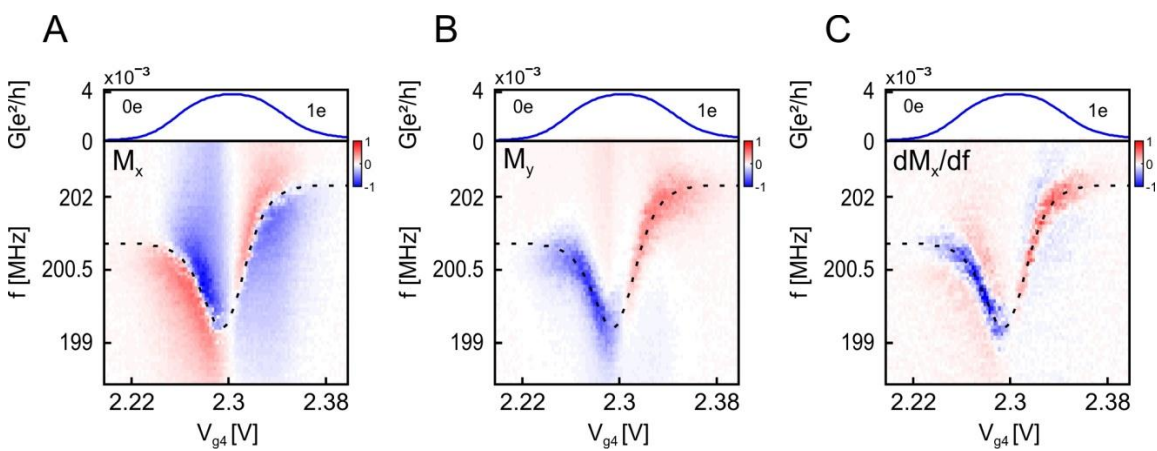

**Figure S1: Different mixing quantities measured around a Coulomb blockade transition.** (A) Top inset: The conductance, $G$, of a quantum dot localized above gate 4, measured as a function of the voltage on gate 4, $V_{g4}$, exhibiting a Coulomb blockade peak that corresponds to the transition from zero to one electron occupation in the dot. Main panel: The in-phase quadrature of the mixing signal, $M_x$ (colormap), measured as a function of the same $V_{g4}$ sweep and the drive frequency, $f$. (B) The out-of-phase mixing quadrature, $M_y$, measured over the same window in $V_{g4}$ and $f$. (C) The numerical derivave of $M_x$ with respect to $f$, $\frac{dM_x}{df}$. For the ease of comparison we show the conductance trace as a top inset in all three panels.

## S2. Determining the tunneling rates from conductance measurements

In this section we explain how the tunneling rates through the left and right barriers of a quantum dot, $\Gamma_L$ and $\Gamma_R$, are determined independently from measurements of its gate-dependent conductance.

The electronic configuration for the measurement is shown in Fig. 2a of the main text: A large quantum dot is created over the three central gates (2 to 4) and is populated with holes. The two side gates (1 and 5) control the transparency of the right and left barriers and the central gate (3) is used as a plunger gate for adding holes to the dot. As a function of gate 3 the dot's conductance exhibits Coulomb blockade oscillations, whose peak values reflect the tunneling rates through the barriers. In the single-electron-transistor regime ($k_B T > \Delta_{\text{dot}}$, $\Delta_{\text{dot}}$ is the level spacing), applicable to our measurements, the conductance at a Coulomb blockade peak is independent of temperature and has a simple form (7): $G_{peak} = \frac{e^2}{2\Delta_{\text{dot}}} \frac{\Gamma_L \Gamma_R}{\Gamma_L + \Gamma_R}$. This peak conductance gives the "series addition" of the two barriers but not their independent tunneling rates.

To extract the two tunneling rates separately we use the gate dependence of $G_{peak}$. For a given pair of side gate voltages, $V_{g1}$ and $V_{g5}$, we scan the center gate voltage, $V_{g3}$, through a Coulomb blockade peak and record the conductance at the peak. By repeating this procedure for different pairs of $V_{g1}$ and $V_{g5}$ we get the full two-dimensional dependence, $G_{peak}(V_{g1}, V_{g5})$, which is plotted as a colormap in Fig 2b of the main text.

To disentangle $\Gamma_L$ and $\Gamma_R$ out of the measured $G_{peak}(V_{g1}, V_{g5})$ we exploit the fact that each barrier depends only on a specific linear combination of gate voltages: The left barrier depends mostly on the voltage of the gate beneath it, $V_{g1}$, to a lesser extent on the central gate voltage, $V_{g3}$, and almost negligibly on the right gate voltage, $V_{g5}$. In fact, using an independent measurement with localized quantum dots (8) we can determine these couplings quantitatively. These coupling are given by the capacitances between the various gates and the NT segment above gate 1, where the barrier is formed ($C_{11}$, $C_{31}$, $C_{51}$). The left barrier is thus affected only through the following combination of gate voltages:

$$\Gamma_L = \Gamma_L(V_L), \qquad V_L = V_{g1} + C_{31} V_{g3}/C_{11} + C_{51} V_{g5}/C_{11} \qquad (S2)$$

Similarly, the right barrier is affected by a different linear-combination of gate voltages:

$$\Gamma_R = \Gamma_R(V_R), \qquad V_R = V_{g5} + C_{35} V_{g3}/C_{55} + C_{15} V_{g1}/C_{55} \qquad (S3)$$

The above relations demonstrate that the two-dimensional dependence $G_{peak}(V_{g1}, V_{g5})$ is in fact simply described only by the one-dimensional functions $\Gamma_L(V_L)$ and $\Gamma_R(V_R)$, and is equal to:

$$G_{peak} = \frac{e^2}{2\Delta_{\text{dot}}} \frac{\Gamma_L(V_L)\Gamma_R(V_R)}{\Gamma_L(V_L) + \Gamma_R(V_R)} \qquad (S4)$$

Figure S2 shows the fit of equations (S2) – (S4) to the measured $G_{peak}(V_{g1}, V_{g5})$. The dashed black lines corresponds to lines of constant $V_L$ and constant $V_R$. The symmetry around the bottom-left to top-right diagonal visible in this figure reflects the symmetry of the barriers, $\Gamma_R = \Gamma_L$, along this line. Specifically, at the bottom-left corner these rates are related to the conductance through: $\Gamma_R = \Gamma_L = \Gamma_0 = G_{peak} 4\Delta_{\text{dot}}/e^2$. Along the line of constant $V_R$ the tunneling rate of the right barrier remains constant ($\Gamma_R = \Gamma_0$) allowing us to extract from equations (S2) and (S4) the dependence of $\Gamma_L$ on $V_L$: $\Gamma_L(V_L) = \left[\frac{e^2}{2\Delta_{dot}G_{peak}} - \frac{1}{\Gamma_0}\right]^{-1}$. Similarly, along the line of constant $V_L$ we extract the voltage-dependence of the right barrier, $\Gamma_R(V_R)$. These two functions, together with equations (S2) – (S4), give the tunneling rates over the entire two-dimensional plane.

The above analysis also gives a critical way to check the validity of the assumptions going into equations (S2) – (S4). To determine the rates we have used the measured conductance only along two lines (dashed black), but now using equations (S2) – (S4) we can predict the conductance over the entire $(V_{g1}, V_{g5})$ two-dimensional plane and compare it with the measured conductance. This comparison is visible in figure S2 overlaying the contours of constant conductance from the measured data (red) and the one calculated by equations (S2) – (S4) (blue). The two exhibit excellent agreement over the entire plane. The small remaining errors, $\Gamma_{err} \sim 10 MHz$, are taken as the horizontal error-bars of the points in Fig 2D in the main text.

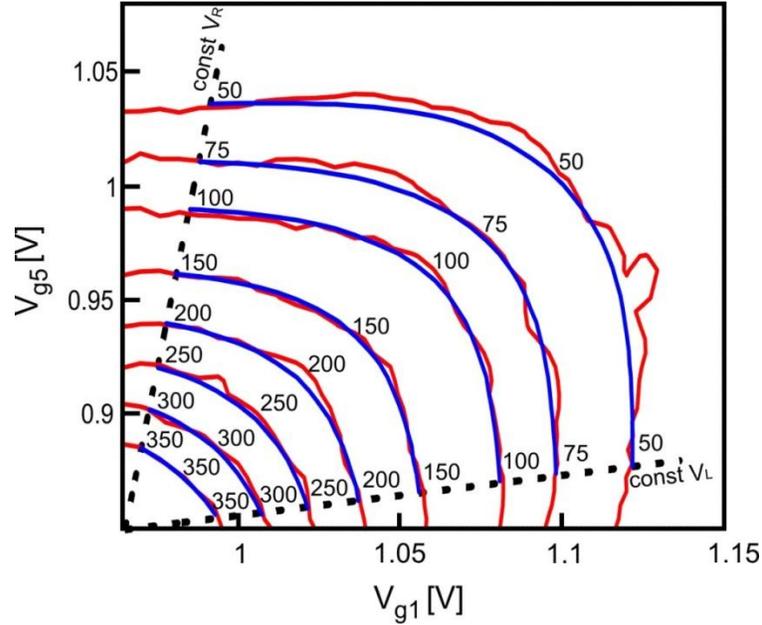

**Figure S2: Measured conductance vs. the prediction of equations (S2) – (S4).** Red contours are the lines of constant conductance taken from the measured $G_{peak}(V_{g1}, V_{g5})$ in Fig. 2B. Blue contours are the line of constant conductance calculated from equations (S2) – (S4) (see text). The numbers show the conductance of each contour, represented as a tunneling rate in MHz. Dashed black lines are the trajectories along which the biasing of the left barrier, $V_L$, or the biasing of the right barrier, $V_R$, are constant (see text).

The conversion of $G$ to $\Gamma$'s requires one parameter, $\Delta_{dot}$, the level spacing in the dot. The latter is obtained from the relation: $\Delta_{dot} = \frac{\hbar v_F}{4}\frac{\pi}{L}$, where $v_F$ is the Fermi velocity, $e$ the electronic charge, $h$ the Plank constant, and $L = 450 nm$ is the length of the 3-gates dot, taken from the lithographic sizes of the gates.

We note that the parameter $\Delta_{dot}$ cancels out in the comparison between theory and experiment, since it appears in a similar manner in the conversion from $G$ to $\Gamma$ (equation

(S4)) and in the theoretical expression for the Γ-dependence of the softening (equation (S20)). As a result we can derive an expression for the softening that is independent of the dot's parameters (equation (S25)). In figure 2 of the main text we obtain a good fit to the measured softening using this expression with T=25K. Compared to the extracted electron temperature of T=16K we see that the quantitative predictions of the simple theory outlined here describes the measured softening reasonably well.

S3.  **The fate of softening at large tunneling rates: the open quantum dot regime.**

In Fig. 2D of the main text we showed the dependence of the mechanical frequency softening, $\Delta f$, on the electron tunneling rate, $\Gamma$, for the regime of a closed quantum dot. In this section we present measurements in the opposite regime, of a very open quantum dot, occurring for large $\Gamma$'s, which exceed the level spacing, $\Gamma \gg \Delta_{\text{dot}}$.

The electronic configuration for this measurement (Fig S3a) is similar to that described in the previous section: A single dot is extended over gates 2-4. The center gate (gate 3) adds holes to the dot and the side gates (1 and 5) control the left and right barriers. Here, however, we 'chain' together gates 1, 3, and 5 such that their voltages change together with the following relation: $\Delta V_{g1} = \Delta V_{g5} = 0.6 \Delta V_{g3}$. Due to this 'chaining' the barriers of the dot are gradually opened concomitant with the gradual addition of holes. Plotting the conductance measured along such a gate voltage sweep (Fig. S3b, top inset) we see that the system evolves continuously from a closed quantum dot with well-developed Coulomb valleys (observed for few holes in the dot, right side), to an open quantum dot with weak conductance modulations (observed for a large number of holes, left side). The main panel of Fig. S3b shows the corresponding mixing signal, $M_x$ (colormap), measured as a function of the same gate voltage sweep and the drive frequency. Interestingly we see that $\Delta f$ is continuously reduced as the dot is gradually opened, in a nice correlation with the smearing of the Coulomb blockade oscillations.

The quenching of $\Delta f$ at large $\Gamma$'s results from the smearing of the charge transitions in the open quantum dot. In an open dot, $\Gamma \gg \Delta_{\text{dot}}$, and the wavefunctions of electrons in the dot "leak" into the leads, reducing the abruptness of charge transitions as compared to the Coulomb blockade regime. Since $\Delta f$ is directly proportional to how quickly the dot's charge changes with gate voltage, $dq_d/dV_g$, it is gradually reduced to the classical-wire softening value as the dot approaches the completely open limit (*9*), in which charge enters continuously with gate voltage. For a completely open dot one thus remains with a frequency softening that corresponds to that of a classical metallic wire, which is much smaller than that observed in the Coulomb blockaded regime.

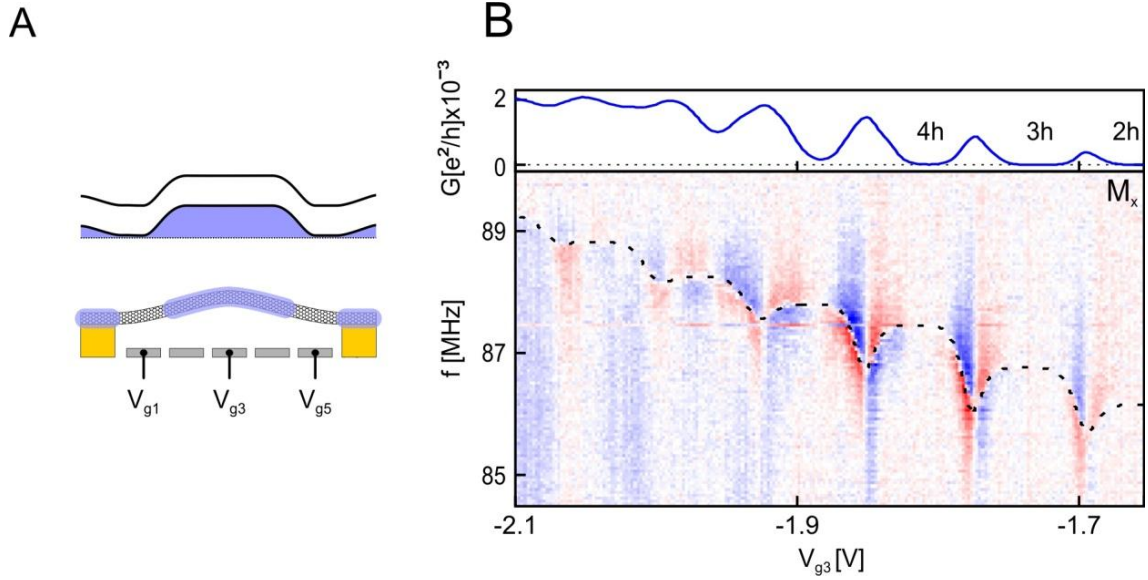

**Figure S3: Measurement of the mechanical softening in the open quantum dot regime.** (A) Illustration of the energy bands and a schematic of the measurement configuration. A dot of holes is defined above gates 2-4. Together with the addition of holes to the dot with a voltage on the center gate, $V_{g3}$, we also change the barrier voltages, $V_{g1}$ and $V_{g5}$, such that with increasing number of holes the dot opens up gradually (see text). (B) Top panel: Conductance $G$, measured as a function of this gate voltage sweep. Bottom panel: Mixing signal, $M_x$ (colormap), measured as a function of this gate voltage sweep and the drive frequency $f$. The dashed black line shows the theoretical fit (see section S1).

Figure S4 summarizes in a schematic plot the different regimes of softening as a function of the tunneling rate. Fundamentally, the softening is a product of two factors: An electrostatic factor, accounting for how abrupt the dot charge changes with gate voltage $dq_{dot}/dV_g$, and a dynamical factor capturing how quick the electrons respond on the vibrational time scales. At zero temperature this product leads to a function that is peaked close to the resonance frequency. At a finite temperature, on the other hand, one obtains an intermediate regime in which the softening is independent of Γ. The characteristic dependence is shown in the figure for the case of $T > \Delta_{dot}$, applicable to our experiments. Noticeably, the softening is quenched both for small rates, $\Gamma \lesssim 2\pi f$, because of slow electron dynamics, and for large rates, $\Gamma \gtrsim \Delta_{dot}$, because of the smearing of the charge transitions. In between ($2\pi f \lesssim \Gamma \lesssim \Delta_{dot}$) the electrons are quick enough to respond, but not too quick to widen the Coulomb blockade peaks beyond their temperature-dominated width, yielding a Γ-independent softening regime. The high temperature in our experiments ($T \approx 16K$) is advantageous in this sense, yielding three order of magnitudes, $100MHz \ll \Gamma \ll 200GHz$, over which the softening is Γ-independent. In the main text we exploited this insensitivity to accurately test the dependence of the softening on other parameters, such as the spatial position (Fig 3).

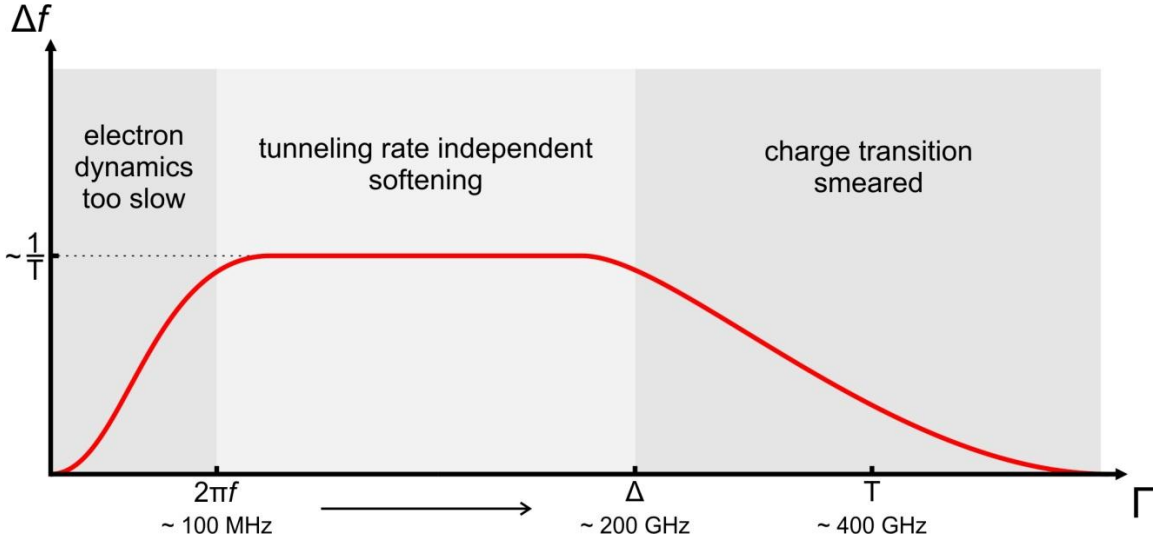

**Figure S4: Schematic dependence of the magnitude of softening, $\Delta f$, on the electrons' tunneling rate, $\Gamma$.** Three regimes are indicated: For $\Gamma \ll 2\pi f$ the softening is quenched because of electron dynamics being too slow, for $\Gamma \gg \Delta_{dot}$ ($\Delta_{dot}$ is the level spacing) it is also quenched because of the smearing of the Coulomb blockade charge transitions. In between, there is a large range of $\Gamma$'s (three orders of magnitude for our experiment) in which softening is independent of $\Gamma$.

S4. **Comparison of the position-dependent softening and vibration mode profiles**

In Fig. 3 we claim that the position dependence of the softening is predominantly due to the position dependence of the vibrational mode amplitude. Since the measured position dependence relies on quantum dots formed at different locations, we should make sure that their parameters do not change as a function of location to give spurious position dependence. In this section we discuss these quantitative aspects of the measurement and its comparison to the calculated profiles. Specifically we demonstrate that the variance in the relevant dot parameters is negligible, that the smearing introduced in the measured profile by the finite size of the dot is insignificant, and finally describe the details of the calculated mechanical mode profiles, to which we compared the measurements.

*I. Position dependence of the dot parameters that are relevant for softening*

Ideally, if exactly similar dots are created at different locations then their effect on the mechanics is to introduce a local 'electronic' spring that is independent of the dot's position (section S8). The frequency shift of the combined mechanical and electronic system, however, will depend on the position of this 'electronic spring' in direct proportionality to the amplitude squared of the displacement profile of the mechanical mode (section S8). Spatial variability of the parameters of the dot or of its electrostatic environment, however, could introduce spurious position dependence. Examining the theoretical expression for the strength of the softening (equation (S29)) we can identify two factors that depend on the dot's parameters: The first is a dynamical factor that depends on the tunneling barriers of the dot. In the previous section (S3) we discussed this effect in detail and showed that in our finite-temperature experiments there is a wide range of Γ's for which the softening is independent of Γ. In the measurements of Fig. 3 we indeed chose the Γ's of all the dots to be well within this range. The second factor is due to the electrostatic environment, amounting to $\left(\frac{e \sum_j C'_{ij} V_j}{C_\Sigma}\right)^2$, where $C'_{ij}$ is the derivative of the capacitive coupling between a gate $j$ and a dot localized above gate $i$, with respect to NT movement in the $z$ direction, and $C_\Sigma$ is the dot's self capacitance. In fact, we can determine all these components with good accuracy: $C_\Sigma$ is measured directly from the Coulomb blockade diamonds. The entire capacitance matrix, $C_{ij}$, is determined directly from gate-dependent transport measurements of localized dots (8). All the elements of this matrix agree quantitatively with finite element simulations of the device (8). Using these simulations we calculate also the z-dependence of all the capacitance elements, $C'_{ij}$. Combined with the gate voltages, $V_j$, these fully determine the softening prefactor. In Fig. 3, in addition to using the same electronic transition in the dots formed at the various locations, we also made sure that their electrostatics, coming through the above prefactor is essentially identical. The effect of the remaining small deviations lead to insignificant changes, as is shown in Fig S5A.

*II. The effect of the finite size of the dot "detector"*

Another effect that may influence the measured position dependence is the finite size of the quantum dots, which effectively 'smears' the measured profile. The magnitude of this effect is estimated in Fig. S5B by convolving the expected profile (solid line) with a Gaussian whose width is equal to the size of the dot (dashed line shows the convoluted profile). Both curves are normalized to match the measured softening above the central gate. As is evident from the comparison, the effect of the smearing is insignificant.

*III. Details of the calculations of the vibrational mode profiles*

The profiles of the mechanical modes in a NT resonator are described by the following differential equation (*10*):

$$EI \cdot z'''' - T \cdot z'' = \mu \omega^2 \cdot z \qquad (S5)$$

Here $E$ is the Young's modulus of the NT, $I = \frac{\pi d^4}{64}$ is its moment of inertia, $d$ is its diameter, $T$ is the tension, $\mu$ is its linear mass density, $z(x)$ is the displacement perpendicular its axis, $\omega$ is the mode frequency, and the primes are derivative with respect to $x$, the coordinate along its axis. This equation has effectively two parameters which we determine by requiring that the first two eigenmodes match the measured resonance frequencies, $\omega_1 = 78MHz$ and $\omega_2 = 197MHz$. The resulting parameters correspond well to realistic values of the physical parameters ($d = 2.15nm$ for the NT diameter, $L = 880nm$ is its suspended length, $E \cdot I = 1.05 TPa \cdot nm^4$ and $\mu = 5.57 \frac{ag}{\mu m}$)

For simplicity, we presented in figures 3K and 3L of the main text the profiles calculated in the string limit ($EI = 0$) and the beam limit ($T = 0$), which are the two extreme cases. Our device is closer to the latter, as can be seen, for example, from the fact that the ratio of its resonance frequencies $\omega_2/\omega_1 = 2.52$, is closer to that in the beam limit ($\omega_2/\omega_1 = 2.76$) than to that in the string limit ($\omega_2/\omega_1 = 2$). As can be seen in figure 3 the differences between the profiles in the two limits is not significant and they both agree well with the measured position-dependent softening.

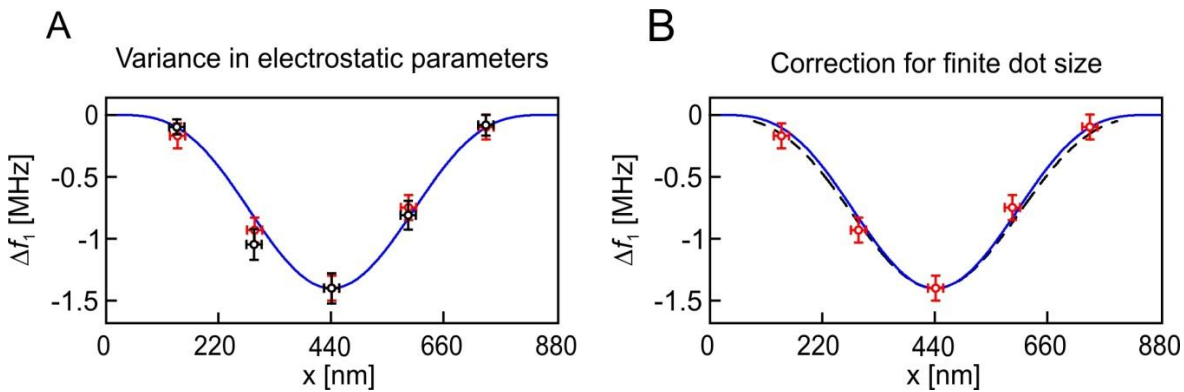

**Figure S5: Quantitative effects on the phonon mode shape imaging.** (A) Effect of position dependence of the dot parameters. Red dots show the measured softening values at the five dot positions (similar to fig 3 in the main text), black dots show the correction of these values for small variance in the dot parameters (see text). Solid line is the phonon profile calculated in the beam limit. (B) The effect of the finite size of the quantum dot "detector". Red dots are the measured softening. Solid line is the profile in the beam limit. Dashed line is its convolution with a 150nm-wide gausian, representing an upper bound on the size of the quantum dots. Both graphs are normalized such that their peak matches the measurement above the central gate.

## S5. Details of the mechanical softening for a double dot

In figure 4G of the main text we present the measured mechanical frequency softening in a double dot configuration as a function of the tunneling rate to the leads. The two dots are defined above gates 2 and 4, and gates 1,3 and 5 control the tunneling rates through the left, center and right barriers, $\Gamma_L$, $\Gamma_C$ and $\Gamma_R$. To determine $\Gamma_L$ and $\Gamma_R$ we work in a regime where the center barrier has a negligible effect on the conductance ($\Gamma_C \gg \Gamma_L, \Gamma_R$) and the right and left barriers are symmetric, $\Gamma_L \approx \Gamma_R$. Then, similar to the procedure described in section S2 we can determine the left and right tunneling rates from the conductance. We set the central barrier tunneling rate to be between $2\pi f_i \ll \Gamma_C \ll \Delta_{dot}$. Under these conditions the internal tunneling dynamics is fast enough to bring the internal tunneling electron to a steady state on the vibrational time scales, but not too fast as to smear its corresponding Coulomb blockade transition line. These considerations on the rate of the center barrier are similar to the ones we explained for the side barriers in figure S4.

The important observation captured in Fig. 4G is that the softening of the 2$^{nd}$ mode remains unchanged when the double-dot is isolated from the leads. To display this in more details we present in figure S6 the measurements of the 2$^{nd}$ mode softening for the two extreme points in Fig 4G: The right inset plots the measured mixing signal, $M_x$ (colormap), as a function of the detuning and frequency, corresponding to the measurement with the largest tunneling to the leads, $\gamma = (\Gamma_L + \Gamma_R)/2\pi f_2$. The left inset show the measurement corresponding to the smallest tunneling to the leads (smallest $\gamma$). Notably, the softening in the isolated case is as large as in the open case, demonstrating the coupling of the 2nd phonon to the internal charge transfer mode in the double-dot.

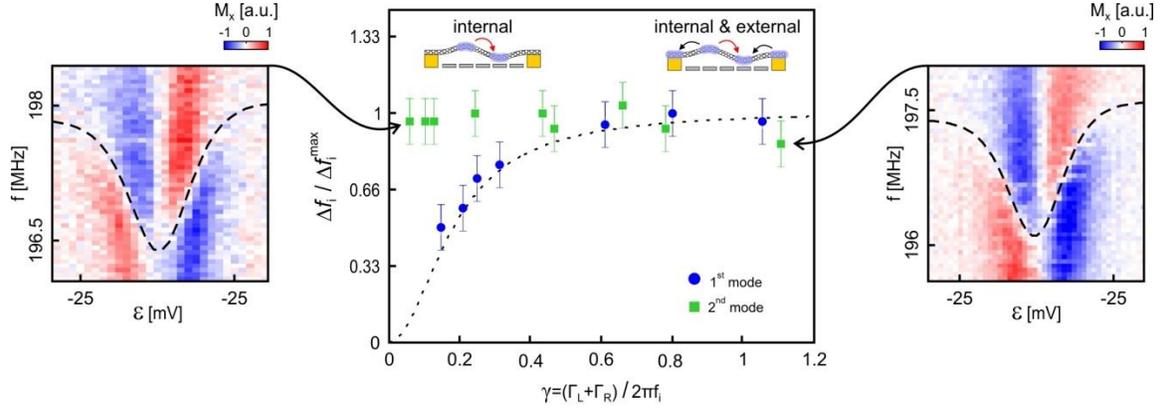

**Figure S6: The softening of the 2nd mechanical mode for a double dot isolated and connected to the leads.** Main panel (reproduced from Fig. 4G in the main text): The measured softening vs. the normalized tunneling rate to the leads, $\gamma = (\Gamma_L + \Gamma_R)/2\pi f_2$. Side inset: the measured mixing signal, $M_x$, corresponding to the point with the largest $\gamma$ (well connected to the leads) and to the point with the smallest $\gamma$ (well isolated from the leads).

## S6. Theory: Rate dependence of softening

In this sectin, we derive the theoretical expressions for the mechanical softening of the vibrational modes of the NT due to the capacitive interaction with a single quantum dot. We focus on the single-electron-transistor limit, $T > \Delta_{dot} > \Gamma$, where $T$ is the temperature, $\Delta_{dot}$ is the level spacing in the dot, and $\Gamma = \Gamma_L + \Gamma_R$ is the combined tunneling rate to the left and right leads.

Following Refs. (*3*, *4*), our starting point is the capacitive energy of a quantum dot capacitively coupled to source (S), drain (D), and gate (g) electrodes,

$$H = -\frac{1}{2}C_g V_g^2 + \frac{(Q_d - C_g V_g)^2}{2 C_\Sigma}. \tag{S6}$$

Here, $Q_d$ denotes the charge of the quantum dot, $C_S, C_D, C_g$ are the capacitances of the quantum dot to the source, drain, and gate electrodes, $C_\Sigma = C_S + C_D + C_g$ is the total capacitance of the dot, and $V_g$ is the applied gate voltage. Inserting this Hamiltonian into the Heisenberg equation of motion $\dot{P} = \frac{i}{\hbar}[H, P]$, where $P$ denotes the momentum associated with the vibrational mode (with displacement variable $z$) and using that the operator $Q_d$ of the quantum dot charge commutes with the mechanical mode operators, we can extract the force induced on the mechanical mode by the electronic subsystem. Since the gate capacitance depends on the mode displacement, $z$, we obtain[1]:

$$F_{elec} = \frac{1}{2}\frac{dC_g}{dz}\left(V_g + \frac{q_d}{C_\Sigma}\right)^2. \tag{S7}$$

Here, we have defined the deviation from the classical gate-induced "control" charge, $Q_C = C_g V_g$, as $q_d = Q_d - Q_C$. Note that $q_d$ jumps by the electron charge, $e$, across a Coulomb blockade peak and varies linearly with gate voltage between Coulomb blockade peaks. For a given gate voltage, the equilibrium displacement of the NT, $z_0(V_g)$, is set by the balance of the electronic force and mechanical restoring force: $F_{elec}(z_0, V_g) + F_{mech}(z_0) = 0$.

Near the Coulomb peak there are two fundametal shifts of the the base resonance frequecy, $\omega_0$:

$$\omega = \omega_0 + \Delta\omega_{static} + \Delta\omega_{dynamic} \tag{S8}$$

The first, static shift, corresponds to a step-like increase in the resonance frequency when going across the Coulomb peak (see, e.g. Fig. 1D). When a single electron is added to the tube the force pulling it toward the gate increases, shifting the NT equilibrium position. If the mechanics of the NT is non linear (deviating from a simple Hooke's law) this will result in a step-like stiffening of its resonance frequency. In this paper we do not focus on this term, which is anyhow absent in the linear regime, but focus on the dynamical frequency shift, which can be very large even in the perfectly linear regime.

The dynamical frequency shift results from the correlated motion of the NT vibrations and the tunneling of electrons in and out of the dot. In the simplest limit (see Refs. (*3*, *4*)), where the frequency of the nanotube resonance mode is small compared to the electronic tunneling rates, the quantum dot charge $q_d$ can be computed for fixed nanotube displacement, $q_d = q_d(z)$, yielding:

$$\frac{\Delta\omega_{dynamic}}{\omega_0} = -\frac{1}{2k_{mech}}\frac{V_g}{C_\Sigma}\frac{dC_g}{dz}\frac{dq_d}{dz}\bigg|_{z_0,\omega_0\to 0} \quad \text{(slow phonon limit),} \tag{S9}$$

where $k_{mech}$ is the elastic spring constant of the vibrational mode of the NT, and small terms proptional to $\frac{d^2C_g}{dz^2}$, $\left(\frac{dC}{dz}\right)^2$ and $q_d^2$ are neglected.

In this section we will go beyond the slow-phonon limit by accounting for the finite rate of electron tunneling (compare Ref. (*11*)).

To obtain the dependence of $q_d$ on mode displacement, we start with the rate equations

---

[1] In general, $C_S$ and $C_D$ may also depend on the mode displacement. We assume that this dependence is weak compared to that of the gate capacitance.

$$\begin{aligned}\frac{dP_N}{dt} &= -\Gamma_+ P_N + \Gamma_- P_{N+1} \\ \frac{dP_{N+1}}{dt} &= -\Gamma_- P_{N+1} + \Gamma_+ P_N\end{aligned}, \quad (S10)$$

where $P_N$ denotes the probability that there are $N$ electrons in the dot. We assume that near the Coulomb blockade peak, only the states with $N$ and $N+1$ electrons on the quantum dot have a finite occupation probability so that $P_N + P_{N+1} = 1$. The average charge on the quantum dot is given by $Q_d = eNP_N + e(N+1)P_{N+1} = eN + eP_{N+1}$. The quantities $\Gamma_+$ and $\Gamma_-$ denote the rates of tunneling into and out of the dot. As the dot's electrostatics changes with the NT displacement $z$, these rates depend on time through $z$. Equations (S10) can be solved for arbitrary time-dependent rates and initial conditions $P_N(t_0)$ and $P_{N+1}(t_0)$, yielding

$$P_N(t) = e^{-\int_{t_0}^t d\tau (\Gamma_+(\tau)+\Gamma_-(\tau))} \left[ P_N(t_0) + \int_{t_0}^t d\tau\, \Gamma_-(\tau) e^{\int_{t_0}^\tau d\tau'(\Gamma_+(\tau')+\Gamma_-(\tau'))} \right]. \quad (S11)$$

The dependence of the tunneling rates on time is given parametrically through the NT displacement $z(t)$. For small NT displacements, we can approximate $\Gamma_\pm[z(t)] \simeq \Gamma_\pm(z_0) + \Gamma'_\pm(z_0)\Delta z(t)$, with $\Gamma'_\pm = d\Gamma_\pm/dz$ and $\Delta z(t) = z(t) - z_0$. With this approximation, we have

$$P_{N/N+1} = \frac{\Gamma_\mp}{\Gamma_+ + \Gamma_-} \pm \frac{\Gamma'_- \Gamma_+ - \Gamma'_+ \Gamma_-}{\Gamma_+ + \Gamma_-} \int_{-\infty}^t d\tau \Delta z(\tau) e^{-(\Gamma_+(t-\tau)+\Gamma_-(t-\tau))}, \quad (S12)$$

where we chose $P_N(t_0) = 1$ and took the limit $t_0 \to -\infty$. The tunneling rates and their derivatives are evaluated at the equilibrium position $z_0$. Using this solution to obtain the charge $q_d(t)$ of the quantum dot as function of time, we find

$$q_d = \frac{e}{2}\frac{\Gamma_+ - \Gamma_-}{\Gamma_+ + \Gamma_-} - e\frac{\Gamma'_- \Gamma_+ - \Gamma'_+ \Gamma_-}{\Gamma_+ + \Gamma_-} \int_{-\infty}^t d\tau \Delta z(\tau) e^{-(\Gamma_+(t-\tau)+\Gamma_-(t-\tau))}, \quad (S13)$$

where we have used $(N + 1/2)e = C_g V_g$, valid in the vicinity of the Coulomb blockade peak, as well as Eq. (S12) and the definition of $q_d$ in terms of $Q_d$. In order to extract the frequency shift, we Fourier transform $q_d$ to the frequency domain,

$$q_\omega = e\pi\delta(\omega)\frac{\Gamma_+ - \Gamma_-}{\Gamma_+ + \Gamma_-}\bigg|_{z_0} - e\frac{\Gamma'_- \Gamma_+ - \Gamma'_+ \Gamma_-}{\Gamma_+ + \Gamma_-}\frac{\Delta z_\omega}{(\Gamma_+ + \Gamma_-) - i\omega}\bigg|_{z_0}, \quad (S14)$$

and take the real part at the unperturbed frequency $\omega_0$ of the vibrational mode The second term on the right-hand side is linear in $z(\omega)$ and thus causes a shift in the resonance frequency,

$$\frac{\Delta \omega_{dynamic}}{\omega_0} = -\frac{e}{2k_{mech}}\frac{V_g}{C_\Sigma}\frac{dC_g}{dz}\frac{\Gamma'_+ \Gamma_- - \Gamma'_- \Gamma_+}{(\Gamma_+ + \Gamma_-)^2 + \omega_0^2}\bigg|_{z_0}. \quad (S15)$$

We mention in passing that the imaginary part in Eq. (S14) describes the electron-induced dissipation of the vibrational mode.

Fermi's Golden Rule provides explicit expressions for the tunneling rates,

$$\Gamma_{\pm} = \pm \sum_\alpha \frac{\Gamma_\alpha}{\Delta_{dot}} \frac{\Delta U}{e^{\pm\beta\Delta U}-1}, \tag{S16}$$

where $\beta = 1/k_B T$ denotes the inverse temperature, $\Gamma_\alpha = 2\pi|t_\alpha|^2 \nu_\alpha$ the partial widths of the dot levels due to tunnel coupling to the source ($\alpha = S$) or drain ($\alpha = D$) leads, with $t_\alpha$ being the tunneling amplitudes and $\nu_\alpha$ the density of states, and $\Delta U = U(N+1) - U(N) = e/C_\Sigma[(N+1/2)e - C_g V_g]$ denotes the electrostatic energy cost to add an electron to the quantum dot. With

$$\frac{d\Delta U}{dz} = -e \frac{V_g}{C_\Sigma} \frac{dC_g}{dz}, \tag{S17}$$

valid near the Coulomb blockade peak, we obtain the relations

$$\begin{aligned} \Gamma'_+\Gamma_- - \Gamma'_-\Gamma_+ &= \frac{\beta}{4}\left[\frac{(\Gamma_R+\Gamma_L)\Delta U}{\Delta_{dot}\sinh(\beta\Delta U/2)}\right]^2 \frac{eV_g}{C_\Sigma} \frac{dC_g}{dz} \\ \Gamma_+ + \Gamma_- &= \frac{(\Gamma_R+\Gamma_L)\Delta U}{\Delta_{dot}\tanh(\beta\Delta U/2)} \end{aligned}. \tag{S18}$$

Substituting Eq. (S18) as well as the definition of $\Delta U$ into (S15), we obtain the frequency shift for arbitrary mode frequency

$$\begin{aligned} \frac{\Delta\omega_{dynamic}}{\omega_0} &= -\frac{\beta}{8k_{mech}}\left(\frac{eV_g}{C_\Sigma}\frac{dC_g}{dz}\right)^2\bigg|_{z_0} \left[\frac{\Gamma\cdot\Delta U}{\Delta_{dot}\sinh(\beta\Delta U/2)}\right]^2 \times \\ &\times \left[\left(\frac{\Gamma\cdot\Delta U}{\Delta_{dot}\tanh(\beta\Delta U/2)}\right)^2 + \omega_0^2\right]^{-1} \end{aligned} \tag{S19}$$

where $\Gamma = \Gamma_R + \Gamma_L$. Note that the frequency shift depends only on the sum of the tunneling rates, emphasizing that the electrons that contribute to the softening can arrive equally well from both leads.

At the Coulomb blockade peak $\Delta U = 0$, simplifying equation (S19) to:

$$\boxed{\frac{\Delta\omega_{dynamic}}{\omega_0} = -\frac{1}{2k_{mech}k_B T}\left(\frac{eV_g}{C_\Sigma}\frac{dC_g}{dz}\right)^2 \cdot \left(1 + \left(\frac{\Delta_{dot}}{2k_B T}\right)^2 \cdot \left(\frac{\omega_0}{\Gamma}\right)^2\right)^{-1}} \tag{S20}$$

To make the expression above more physically transparent we note that the classically induced charge on the NT, the "control charge", has an electronic and mechanical contributions:

$$\delta Q_c = \delta Q_c^{elec} + \delta Q_c^{mech} \equiv C_g \delta V_g + V_g \delta C_g \tag{S21}$$

The mechanical component of the control charge results from the gate capacitance change due to NT movement, which is also induced by the gate voltage:

$$\delta Q_c^{mech} \equiv V_g \frac{dC_g}{dz}\frac{dz}{dV_g}\delta V_g = \frac{1}{k_{mech}}\left(V_g \frac{dC_g}{dz}\right)^2 \delta V_g, \tag{S22}$$

In the last equation we calculated $\frac{dz}{dV_g}$ from a balance between a linear mechanical force and the classical electrostatic force, $k_{mech}z = \frac{1}{2}\frac{dC_g}{dz}V_g^2$. Defining also the charging energy, $E_c = \frac{e^2}{C_\Sigma}$, and the leverarm factor, $\alpha = \frac{C_g}{C_\Sigma}$, we can recast the dynamic coupling frequency shift in a simple form:

$$\boxed{\frac{\Delta\omega_{dynamic}}{\omega_0} = -\frac{1}{2}\alpha\left(\frac{\delta Q_C^{mech}}{\delta Q_C^{elec}}\right)\cdot\left(\frac{E_c}{k_BT}\right)\cdot\left(1+\left(\frac{\Delta_{dot}}{2k_BT}\right)^2\cdot\left(\frac{\omega_0}{\Gamma}\right)^2\right)^{-1}} \quad (S23)$$

The first term is the electrostatic leverarm factor. The second term $\left(\frac{\delta Q_C^{mech}}{\delta Q_C^{elec}}\right)$ could be refered to as the "mechanical leverarm factor" – it describes the ratio between the charge induced on the tube due to its mechanical motion and to that induced by electrostatics. The third factor describes the Coulomb blockade enhancement of the rate of charge addition with gate voltage. In comparrison to a classical wire, where charge is continuously added with gate voltage, in the Coulomb blockade regime the charge enters only around the Coulomb peak. The rate of charge addition is therefore enhanced by the ratio of the peak spacing (charging energy) and the peak width (temperature, in the single-electron transistor limit), $\left(\frac{E_c}{k_BT}\right)$. The last factor, $\left(1+\left(\frac{\Delta_{dot}}{2k_BT}\right)^2\cdot\left(\frac{\omega_0}{\Gamma}\right)^2\right)^{-1}$, captures the relative dynamics of the electrons and vibrations and is sensitive to the ratio of their frequencies.

Equation (S23) includes one parameter of the dot, its level spacing $\Delta_{dot}$, which is not always measured directly. Interestingly, however, the conversion of the dot's peak conductance to the tunneling rates also includes $\Delta_{dot}$ in a way that leads to a perfect cancelation of this parameter. For symmetric barriers, $\Gamma_R = \Gamma_L$, the conductance at the Coulomb peak is

$$G_{peak} = \frac{e^2}{2\Delta_{dot}}\frac{\Gamma_L\Gamma_R}{\Gamma_L+\Gamma_R} = \frac{e^2}{8\Delta_{dot}}\Gamma. \quad (S24)$$

Substituting this in equation (S23), and defining the unitless conductance as $g_{peak} = G_{peak}/(e^2/h)$, we get

$$\frac{\Delta\omega_{dynamic}}{\omega_0} = --\frac{1}{2}\alpha\left(\frac{\delta Q_C^{mech}}{\delta Q_C^{elec}}\right)\cdot\left(\frac{E_c}{k_BT}\right)\cdot\left(1+\left(\frac{1}{16}\cdot\frac{\hbar\omega_0}{k_BT}\cdot\frac{1}{g_{peak}}\right)^2\right)^{-1}, \quad (S25)$$

where the roll-off at low tunneling rates is described in terms of the thermal occupation of the vibrational mode, $\frac{k_BT}{\hbar\omega_0}$, and the unitless conductance, $g_{peak}$.

## S7. Theory: Generalization to multiple gates

In the previous section we assumed, for the simplicity of the derivation, that the quantum dot is coupled to a single gate. In this section we generalize this for a multi-gated geometry and show that a similarly-simple expression holds.

The generalization of equation (S6) to multiple gates, ignoring the 'classical wire' contributions gives:

$$H = \frac{1}{2C_\Sigma}\left(Q_i - \sum_j C_{ij}V_j\right)^2. \tag{S26}$$

Here $C_{ij}$ is the capacitive coupling between gate $j$ and the NT segment above gate $i$, which we measure directly (8). The Hamiltonian assumes that a quantum dot is formed locally at the NT segment above gate $i$ and its charge is $Q_i$.

In analogy to the previous section, we define the gates-induced "control charge" on the dot as $Q_i^c = \sum_j C_{ij}V_j$, the actual charge in the dot as $Q_i^d$, their difference as $q_d = Q_i^d - Q_i^c$, the total capacitance of the dot as $C_\Sigma = C_S + C_D + \sum_j C_{ij}$, and an effective gate voltage given by $V_{eff} = Q_i^{c\prime}/C_\Sigma'$, where the primes correspond to a derivative with respect to $z$. With these relations the Hamiltonian becomes:

$$H = \frac{1}{2C_\Sigma}\left(Q_i^d - Q_i^c\right)^2 \tag{S27}$$

And the corresponding force and softening, in analogy to equations (S7) and (S20), are:

$$F = -H' = \frac{1}{2}C_\Sigma'\left(\frac{q_d}{C_\Sigma} + V_{eff}\right)^2 \tag{S28}$$

$$\frac{\Delta\omega_{dynamic}}{\omega_0} = -\frac{1}{2k_{mech}k_BT}\left(\frac{e\sum_j C_{ij}'V_j}{C_\Sigma}\right)^2 \cdot \left(1 + \left(\frac{\Delta_{dot}}{2k_BT}\right)^2 \cdot \left(\frac{\omega_0}{\Gamma}\right)^2\right)^{-1} \tag{S29}$$

where small terms proptional to $\frac{d^2C_g}{dz^2}$, $\left(\frac{dC}{dz}\right)^2$ and $q_d^2$ were neglected.

## S8. Theory: Spatial dependence of the softening

In this section we explain the connection between the spatial dependence of the softening and the profile of the vibrational mode along the NT axis. We show that the effect of a local quantum dot can be mapped onto an electronic 'spring', with a negative spring constant connected at the position of the dot (Fig S7). This spring leads to a softening that is proportional to the amplitude squared of the bare vibrational mode at its connection point.

The electrostatic force acting on a quantum dot localized on the NT is given by equation (S7). This force depends on the height of the dot with respect to the gates, $z(x)$, and thus can be mapped into an effective spring, connected at the position of the dot, $x_0$, perpendicular to the NT axis. The Hooke's constant of this 'electronic' spring is:

$$k_{elec} = -dF_{elec}/dz \tag{S30}$$

This electronic spring adds to the NT's elastic forces, altering its resonance frequency. If the electrostatic environment of the dot and the tunneling rates of its barriers are independent of $x_0$ (as is demonstrated experimentally in section S4), the electronic spring constant, $k_{elec}$, does not dependent on the dot's position. but its effect on the frequency of the combined system does. If the spring is connected at a position where the bare vibration mode has a large amplitude its effect will be large, and if it is connected at a node it would have no effect.

To derive the position dependence quantitatively, we assume that the electronic contribution is small and consider it perturbatively. In the absence of the perturbation the profile of any eigenmode of the beam, $z_0(x) = Z \cdot \zeta_0(x)$ (where Z is the overall amplitude and $\zeta_0(x)$ is the unit-less mode shape), and its eigenfrequency, $\omega_0$, are given by equation (S5), which in an operator form reads:

$$\widehat{D} z_0(x) = \mu \omega_0^2 z_0(x), \tag{S31}$$

where $\widehat{D} = EI \cdot \frac{\partial^4}{\partial x^4} - T \cdot \frac{\partial^2}{\partial x^2}$ and $x$ is the coordinate along the NT. In the lowest order in the local electronic perturbation, $k_{elec}\delta(x - x_0)$, the eigenmode remains the same, but the eigenfrequency changes to $\omega$, given by:

$$\left(\widehat{D} + k_{elec}\delta(x - x_0)\right) \cdot z_0(x) = \mu \omega^2 z_0(x) \tag{S32}$$

Using equation (S31) we get:

$$k_{elec}\delta(x - x_0) \cdot z_0(x) = \mu(\omega^2 - \omega_0^2) z_0(x) \tag{S33}$$

Multiplying both sides by $z_0(x)$, and using the fact that the bare mode is normalized over the length of the beam, $\frac{1}{L}\int_0^L \zeta_0(x)^2 dx = 1$, we get after integration that:

$$\frac{k_{elec}}{L}\zeta_0^2(x_0) = \mu(\omega^2 - \omega_0^2) \tag{S34}$$

Defining $\omega = \omega_0 + \delta\omega$, we get that to the lowest order in the electronic perturbation the frequency shift is directly proportional to the amplitude squared of the bare mode at the position of the local dot:

$$\frac{\delta\omega}{\omega_0} = \frac{k_{elec}}{2k_{mech}} \cdot \zeta_0^2(x_0) \tag{S35}$$

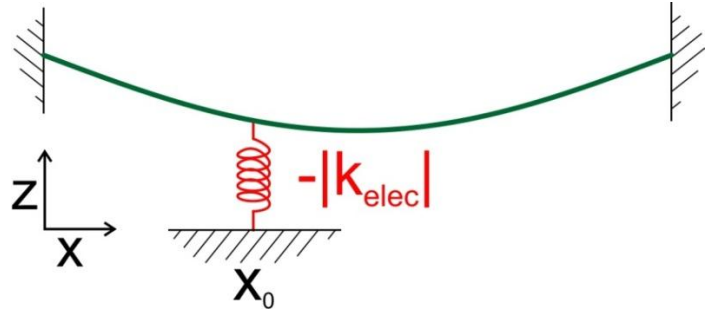

**Figure S7: A equivalent mechanical model for spatial dependence of the softening.** The effect of a local quantum dot at a position $x_0$ along the NT is mapped to an effective 'electronic spring', with a negative spring constant, connected at this position perpendicular to the NT (see text).

### S9. Theory: Softening with a double-dot

For double quantum dots, the electron-phonon coupling becomes much more sensitive to the spatial structure of the vibrational modes of the NT. This produces additional effects in the frequency shifts compared to the single-quantum-dot case. In this section, we focus on the interaction with the two lowest phonon modes although the theoretical considerations can in principle be readily extended to higher vibrational modes. The considerations of the present section form the basis of our analysis related to Fig. 4 of the main text.

*I. Electrostatic coupling*

The electrostatic energy of double quantum dots is given by (*12*)

$$\begin{aligned} H &= -\tfrac{1}{2} C_{g1} V_{g1}^2 - \tfrac{1}{2} C_{g2} V_{g2}^2 \\ &+ \tfrac{1}{\Delta_c} C_{12} (Q_1 - C_{g1} V_{g1})(Q_2 - C_{g2} V_{g2}) \\ &+ \tfrac{1}{2\Delta_c} C_2 (Q_1 - C_{g1} V_{g1})^2 + \tfrac{1}{2\Delta_c} C_1 (Q_2 - C_{g2} V_{g2})^2 . \end{aligned} \quad (S36)$$

Here, we included the capacitive coupling $C_{S/D}$ between left/right dot and source/drain lead, the capacitive couplings $C_{g1}$ and $C_{g2}$ to the left and right gate electrodes with gate voltages $V_{g1}$ and $V_{g2}$, as well as the interdot capacitance $C_{12}$. The charges on the quantum dots are denoted by $Q_1$ and $Q_2$, respectively. We also defined the shorthands $C_{1(2)} = C_{S(D)} + C_{g1(g2)} + C_{12}$ and $\Delta_c = C_1 C_2 - C_{12}^2$. The source-drain bias is set to zero. For simplicity in the following we consider a symmetric configuration in which $C_S = C_D$, hence we set $C_S$ as the capacitance to the leads.

The electron-phonon coupling is due to the dependence of the gate capacitances $C_{gi}$ on the displacements of quantum dots $i = 1,2$. When the quantum dots are placed symmetrically about the center of the NT, these displacements are characteristically different for the first and the second phonon mode: They are equal for the first phonon mode, but have opposite signs and the same magnitude for the second mode. Correspondingly, the variations of the gate capacitances with the effective mode

displacement $\Delta z$ away from the equilibrium displacement $z_0$ can be approximated through

$$\begin{aligned}C_{g1}^{\pm}(z) &= C_g(z_0) + C_g'(z_0)\Delta z \\ C_{g2}^{\pm}(z) &= C_g(z_0) \pm C_g'(z_0)\Delta z,\end{aligned} \quad (S37)$$

where the upper sign (lower sign) refers to the first (second) vibrational mode. Thus, variations of $\Delta z$ effectively modify the classical gate charges in the same way as changes in the gate voltages $V_{g1}$ and $V_{g2}$ along the directions indicated in Figs. 4E.

## II. Frequency-dependent softening

We first consider the frequency shift for the case that there is no tunneling between the leads and the quantum dots while interdot tunneling is finite. In this situation, one expects that the first mode is barely affected by the electron-phonon coupling as the charges on the quantum dots will vary only very weakly with the corresponding phonon displacement. In particular, we do not expect any effect for symmetric gate voltages $V_{g1} = V_{g2}$. In contrast, one expects a much larger effect for the second phonon mode. Here, the asymmetric change in the gate capacitances with $\Delta z$ causes a charge flow between the two quantum dot and hence a significant frequency shift. In this case, the charge flow and the frequency shift are expected to be largest for $V_{g1} = V_{g2}$.

For vanishing tunnel coupling to the leads and for near symmetric gate voltages, the relevant charge configurations of the double quantum dot are $(N, N+1)$ and $(N+1, N)$. Then, the rate equations describing the occupation probabilities of these two states take the form

$$\begin{aligned}\frac{d}{dt}P_{N+1,N} &= -\Gamma_{12}P_{N+1,N} + \Gamma_{21}P_{N,N+1} \\ \frac{d}{dt}P_{N,N+1} &= -\Gamma_{21}P_{N,N+1} + \Gamma_{12}P_{N+1,N}\end{aligned}. \quad (S38)$$

Here, $P_{N,N+1}$ ($P_{N+1,N}$) denotes the probability of the configuration $(N, N+1)$ $((N+1, N))$. $\Gamma_{12}$ ($\Gamma_{21}$) is the tunneling rate of electrons from quantum dot 1 (2) to quantum dot 2 (1), which depends parametrically on time through $z(t)$. The rate equations are formally identical to those in Eq. (S10) for a single quantum dot and hence, we can write down its solution by analogy with Eq. (S12) through the identifications $P_{N+1,N(N,N+1)} \leftrightarrow P_{N(N+1)}$ and $\Gamma_{12(21)} \leftrightarrow \Gamma_{+(-)}$. Thus, we find for the average charges in the quantum dots in frequency space,

$$\begin{aligned}[Q_{1(2)}]_\omega^{inphase} &= e2\pi\delta(\omega)\left(N + \frac{\Gamma_{21(12)}}{\Gamma_{DD}}\right) \\ &\quad -(+)e\frac{\Gamma_{12}'\Gamma_{21}-\Gamma_{21}'\Gamma_{12}}{\Gamma_{DD}^2+\omega^2}\Delta z_\omega,\end{aligned} \quad (S39)$$

where we specified the in-phase charge response to the mechanical motion, which is responsible for the frequency shift. We also introduced the shorthand $\Gamma_{DD} = \Gamma_{12} + \Gamma_{21}$.

By analogy with the calculation for the single quantum dot, we obtain the frequency shifts of the first ($\Delta\omega^+_{dynamic}$) and second ($\Delta\omega^-_{dynamic}$) vibrational modes of the NT. We find

$$\begin{aligned}
\frac{\Delta\omega^+_{dynamic}}{\omega_0} &= \frac{e}{2k_{mech}} C'_g \left( \frac{\Gamma'_{12}\Gamma_{21} - \Gamma'_{21}\Gamma_{12}}{\Gamma^2_{DD} + \omega_0^2} \right) \\
&\times \left[ \frac{2C_{12}(C_g + C_s)}{\Delta_c^2}(\Delta Q - C_g \Delta V) - \frac{1}{\Delta_c}(\Delta Q + C_s \Delta V) \right] \\
\frac{\Delta\omega^-_{dynamic}}{\omega_0} &= \frac{e}{2k_{mech}} C'_g \left( \frac{\Gamma'_{12}\Gamma_{21} - \Gamma'_{21}\Gamma_{12}}{\Gamma^2_{DD} + \omega_0^2} \right) \frac{1}{\Delta_c} (\bar{Q} + C_s \bar{V})
\end{aligned} \quad (S40)$$

with $\Delta Q = \langle Q_2 \rangle - \langle Q_1 \rangle$, $\bar{Q} = \langle Q_2 \rangle + \langle Q_1 \rangle$, $\Delta V = (V_{g2} - V_{g1})$, and $\bar{V} = V_{g1} + V_{g2}$. All quantities are evaluated at the equilibrium position $z_0$. The tunneling rates can be evaluated by Fermi's Golden Rule. Within the constant-interaction model, one obtains

$$\Gamma_{21(12)} = +(-)T_M \frac{\Delta U_{DD}}{e^{\pm \beta \Delta U_{DD}} - 1}, \quad (S41)$$

in analogy with Eq. (S16). Here, $T_M$ is the inter-dot transition rate and $\Delta U_{DD} = U_{DD}(N+1, N) - U_{DD}(N, N+1)$ the electrostatic energy cost for an electron to tunnel from dot 2 to dot 1, where $U_{DD}(N, N')$ is the eigenenergy of the double-dot Hamiltonian (Eq. (S36)) for the charge state $(N, N')$. $\Delta U_{DD}$ can be evaluated to first order in the displacement

$$\begin{aligned}
\Delta U^+_{DD} &= \frac{eC_g}{\Delta_c}(C_g + C_s)\Delta V + \left[\frac{eC_g}{\Delta_c}(C_g + C_s)\right]' \Delta z \Delta V \\
\Delta U^-_{DD} &= \frac{eC_g}{\Delta_c}(C_g + C_s)\Delta V - \frac{eC'_g}{\Delta_c}(\bar{Q} + C_s \bar{V})\Delta z
\end{aligned} \quad (S42)$$

for the first and second vibrational modes, respectively. The relevant quantities that enter Eq. (S40) are

$$\begin{aligned}
\Gamma'_{12}\Gamma_{21} - \Gamma'_{21}\Gamma_{12} &= \frac{\beta}{4} \left[ \frac{T_M \Delta U_{DD}}{\sinh(\beta \Delta U_{DD}/2)} \right]^2 \frac{dU_{DD}}{dz} \\
\Gamma_{DD} &= (T_M) \frac{\Delta U_{DD}}{\tanh(\beta \Delta U_{DD}/2)},
\end{aligned} \quad (S43)$$

with $\Delta U_{DD} = \Delta U^\pm_{DD}$ as given in Eq. (S42). Inserting Eqs. (S39), (S42) and (S43) into Eq. (S40) we obtain the final expressions for the frequency shifts $\Delta\omega^\pm_{dynamic}$ due to inter-dot tunneling. The expressions for the frequency-dependent shifts are lengthy and we will not reproduce them here. It is worth noting however that $\Delta\omega^+_{dynamic} \propto \Delta V$ and hence the shift vanishes along the $V_1 = V_2$ line, while $\Delta\omega^-_{dynamic}$ is finite, in agreement with Fig. 4G.

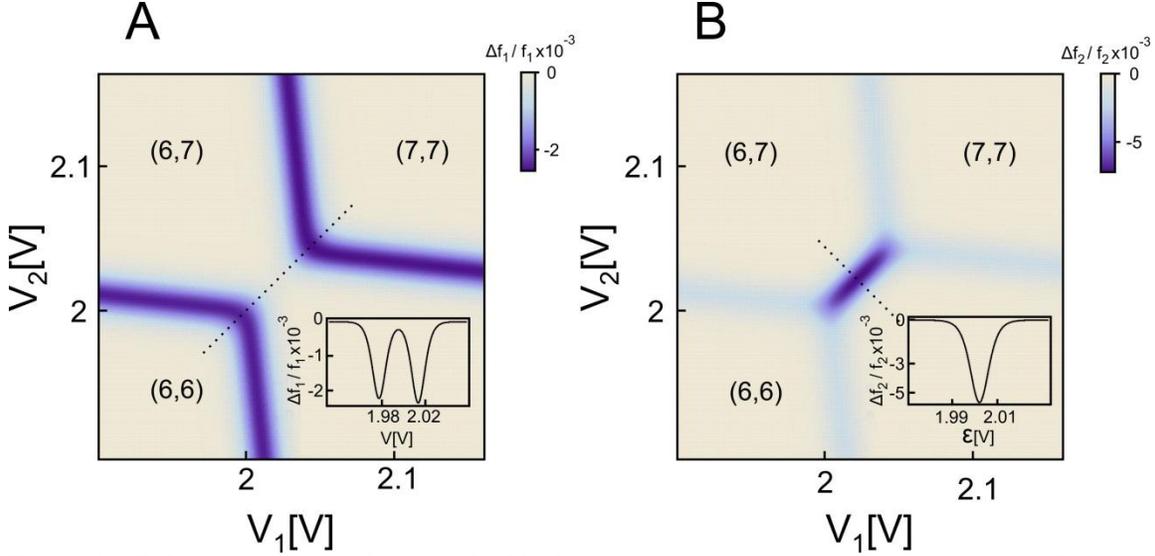

**Figure S8: Relative frequency shifts for double dot configuration for four charge states.** (A) Frequeny shift for the first mode as a function of the voltages $V_{g1}$ and $V_{g2}$. The inset shows the frequency shift along the dashed line. (B) Frequency shift for the second mode. The inset shows the frequency shift along the dashed line. For both calculations we used the parameters that correspond to our experiments: $k_{mech} = 700\mu N/m$, $C_{12} = 0.1$ aF, $C_g = 0.5$ aF, $T = 16$ K, $N = 6$ and $C_S = C_D = 1.1$aF.

This is most easily seen for the limit of small frequencies and $C_S = 0$, in which the general expressions given in Eq. (S40) simplify considerably to

$$\left.\frac{\Delta\omega^+_{dynamic}}{\omega_0}\right|_{\omega_0 \to 0} = -\frac{\beta}{4k_{mech}}\left[\frac{eC'_g}{(C_g+2C_{12})^2\cosh(\beta\Delta U^+_{DD}/2)}\right]^2 \times$$
$$\times C_{12}\Delta V\left[\tanh(\beta\Delta U^+_{DD}/2) + 2\frac{C_{12}\Delta V}{e}\right] \quad \text{(S44)}$$

$$\left.\frac{\Delta\omega^-_{dynamic}}{\omega_0}\right|_{\omega_0 \to 0} = -\frac{e\beta}{8k_{mech}}\left[\frac{eC'_g(2N+1)}{C_g(C_g+2C_{12})\cosh(\beta\Delta U^+_{DD}/2)}\right]^2$$

with

$$\Delta U^+_{DD}|_{C_S=0} = \frac{eC_g}{C_g+2C_{12}}\Delta V + \frac{2eC_{12}C'_g}{(C_g+2C_{12})^2}\Delta z \Delta V$$
$$\Delta U^-_{DD}|_{C_S=0} = \frac{eC_g}{C_g+2C_{12}}\Delta V - \frac{e^2 C'_g(1+2N)}{C_g(C_g+2C_{12})}\Delta z \quad \text{(S45)}$$

where we have used $\bar{Q} = e(1+2N)$ and $\Delta Q = e\tanh(\beta\Delta U_{DD}/2)$.

**Softening including four charge states:** In the previous section we considered the case of vanishing tunneling to the leads, for which the relevant physics is dominated by two charge states. Here we allow for finite tunneling to the leads and hence need to include all four possible charge states, $(N, N+1)$, $(N+1, N)$, $(N, N)$, $(N+1, N+1)$. We study the adiabatic limit, for which the tunneling rates are bigger than the resonant frequency $\omega_0$. These considerations apply to the right hand side of Fig. 4G in the main text.

We start by solving the rate equations in static equilibrium, for which the tunneling rates are time independent. These equations can be written, using Fermi's Golden Rule for every possible tunneling event, as

$$\frac{dP_{N,N}}{dt} = -P_{N,N}[T_L W(N+1,N:N,N) + T_R W(N,N+1:N,N)]$$
$$+ P_{N+1,N} T_L W(N,N:N+1,N) + P_{N,N+1} T_R W(N,N:N,N+1) = 0$$

$$\frac{dP_{N,N+1}}{dt} = -P_{N,N+1}[T_L W(N+1,N+1:N,N+1) + T_R W(N,N:N,N+1) + T_M W(N+1,N:N,N+1)]$$
$$+ P_{N,N} T_R W(N,N+1:N,N) + P_{N+1,N+1} T_L W(N,N+1:N+1,N+1)$$
$$+ P_{N+1,N} T_M W(N,N+1:N+1,N) = 0$$

$$\frac{dP_{N+1,N}}{dt} = -P_{N+1,N}[T_L W(N,N:N+1,N) + T_R W(N+1,N+1:N+1,N) + T_M W(N,N+1:N+1,N)]$$
$$+ P_{N,N} T_L W(N+1,N:N,N) + P_{N+1,N+1} T_R W(N+1,N:N+1,N+1)$$
$$+ P_{N,N+1} T_M W(N+1,N:N,N+1) = 0$$

$$1 = P_{N,N} + P_{N,N+1} + P_{N+1,N} + P_{N+1,N+1},$$

(S46)

where we have defined the weight

$$W(N,N':M,M') = \frac{U_{DD}(N,N') - U_{DD}(M,M')}{e^{\beta[U_{DD}(N,N') - U_{DD}(M,M')]} - 1}.$$

(S47)

The solution to the system of equations (S46) is independent of the transition rates as observed in Fig. 4G of the main text. Explicitly,

$$P_{N,N} = \frac{e^{\beta[U_{DD}(N+1,N+1) + U_{DD}(N,N+1) + U_{DD}(N+1,N)]}}{D}$$

$$P_{N+1,N+1} = \frac{e^{\beta[U_{DD}(N,N) + U_{DD}(N,N+1) + U_{DD}(N+1,N)]}}{D}$$

$$P_{N+1,N} = \frac{e^{\beta[U_{DD}(N,N) + U_{DD}(N+1,N+1) + U_{DD}(N,N+1)]}}{D}$$

$$P_{N,N+1} = \frac{e^{\beta[U_{DD}(N,N) + U_{DD}(N+1,N+1) + U_{DD}(N+1,N)]}}{D}$$

(S48)

with

$$D = e^{\beta[U_{DD}(N,N) + U_{DD}(N,N+1) + U_{DD}(N+1,N)]}$$
$$+ e^{\beta[U_{DD}(N,N) + U_{DD}(N+1,N+1) + U_{DD}(N,N+1)]}$$
$$+ e^{\beta[U_{DD}(N+1,N+1) + U_{DD}(N,N+1) + U_{DD}(N+1,N)]}$$
$$+ e^{\beta[U_{DD}(N,N) + U_{DD}(N+1,N+1) + U_{DD}(N+1,N)]}.$$

(S49)

With Eq. (S46) we can calculate the average charge in each dot as a function of position, and therefore the (static) resonant frequency shift as in the previous sections, both for the first and second vibrational modes of the NT. The results of this procedure lead to the frequency shifts are plotted in Fig. S8.

**References:**

1. V. Sazonova *et al.*, A tunable carbon nanotube electromechanical oscillator., *Nature* **431**, 284–7 (2004).

2. B. Witkamp, M. Poot, H. S. J. van der Zant, Bending-mode vibration of a suspended nanotube resonator., *Nano Letters* **6**, 2904–8 (2006).

3. G. A. Steele *et al.*, Strong coupling between single-electron tunneling and nanomechanical motion., *Science* **325**, 1103–7 (2009).

4. B. Lassagne *et al.*, Coupling mechanics to charge transport in carbon nanotube mechanical resonators., *Science* **325**, 1107–10 (2009).

5. V. Gouttenoire *et al.*, Digital and FM Demodulation of a Doubly Clamped Single-Walled Carbon-Nanotube Oscillator: Towards a Nanotube Cell Phone., *Small* , 1060–1065 (2010).

6. C. C. Wu, Z. Zhong, Capacitive spring softening in single-walled carbon nanotube nanoelectromechanical resonators., *Nano Letters* **11**, 1448–51 (2011).

7. C. W. J. Beenakker, Theory of Coulomb-blockade oscillations in the conductance of a quantum dot, *Physical Review B* **44**, 1646–1656 (1991).

8. J. Waissman *et al.*, Electronically-Pristine and Locally-Tunable One-Dimensional Systems Created in Carbon Nanotubes Using Nano-Assembly, *Preprint at arXiv:1302.2921* .

9. H. B. Meerwaldt *et al.*, Probing the charge of a quantum dot with a nanomechanical resonator, *Physical Review B* **86**, 115454 (2012).

10. V. Sazonova, thesis, Cornell University (2005).

11. C. Lotze, M. Corso, K. J. Franke, F. von Oppen, J. I. Pascual, Driving a Macroscopic Oscillator with the Stochastic Motion of a Hydrogen Molecule, *Science* **338**, 779–782 (2012).

12. W. van der Wiel, S. De Franceschi, Electron transport through double quantum dots, *Reviews of Modern Physics* **75**, 1–22 (2002).